\documentclass[10pt,twocolumn,letterpaper]{article}

\usepackage[pagenumbers]{cvpr}  
\usepackage{bm}

\usepackage{array}
\usepackage{color, colortbl}
\usepackage{xcolor} 
\usepackage{multirow}
\usepackage{graphicx}
\usepackage{placeins}
\usepackage{microtype}

\usepackage{adjustbox,booktabs}

\usepackage[normalem]{ulem} 

\newcommand{\new}[1]{{\color{black} #1}}

\def\HIDEREMOVED{}

\ifdefined\HIDEREMOVED
  \newcommand{\newremove}[1]{}
\else
  \usepackage[normalem]{ulem} 
  \usepackage{xcolor}
  \newcommand{\newremove}[1]{\textcolor{blue}{[\sout{#1}]}}
\fi

\usepackage{amssymb}
\usepackage{graphicx}
\usepackage{calc}
\usepackage{tikz}
\renewcommand{\paragraph}[1]{\vspace{.5em}\noindent\textbf{#1.}}

\definecolor{ourgreen}{RGB}{106, 204, 100}
\definecolor{ourred}{RGB}{238, 133, 34}

\usepackage{pifont}
\newcommand{\cmark}{\ding{51}}%
\newcommand{\xmark}{\ding{55}}%

\newcommand\rurl[1]{%
  \href{https://#1}{\nolinkurl{#1}}%
}
\usepackage[shortcuts]{extdash}

\usepackage[utf8]{inputenc}
\definecolor{cvprblue}{rgb}{0.21,0.49,0.74}
\usepackage[pagebackref,breaklinks,colorlinks,citecolor=cvprblue]{hyperref}

\def\paperID{} 
\def\confName{CVPR}
\def\confYear{2026}

\title{EDGS: Eliminating Densification for Efficient Convergence of 3DGS}

\author{
Dmytro Kotovenko$^*$ \and 
Olga Grebenkova$^*$ \and 
Björn Ommer \and \\
CompVis @ LMU Munich,  \quad Munich Center for Machine Learning (MCML)
}

\begin{document}
\twocolumn[{
\renewcommand\twocolumn[1][]{#1}
    \maketitle
    \begin{center}
    \vspace{-10pt}
        \rurl{compvis.github.io/EDGS}
    \end{center}
     \begin{center}
      \captionsetup{type=figure}
         \includegraphics[width=0.32\linewidth]{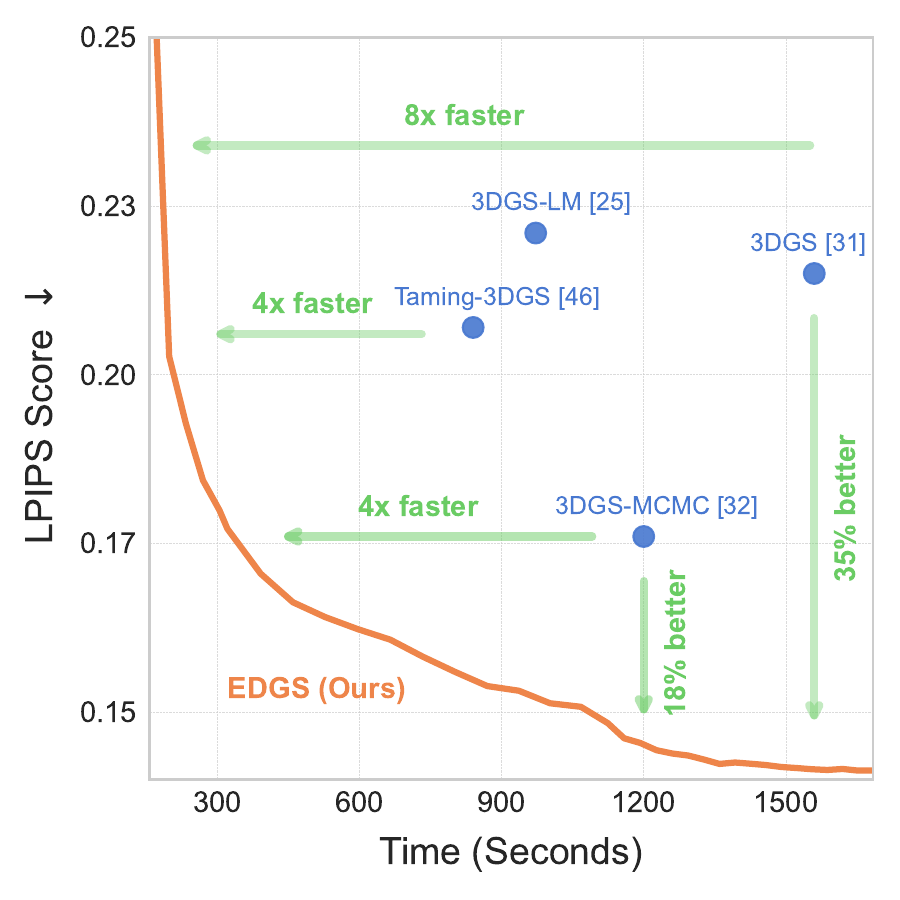}
      \hfill
       \includegraphics[width=.67\linewidth]{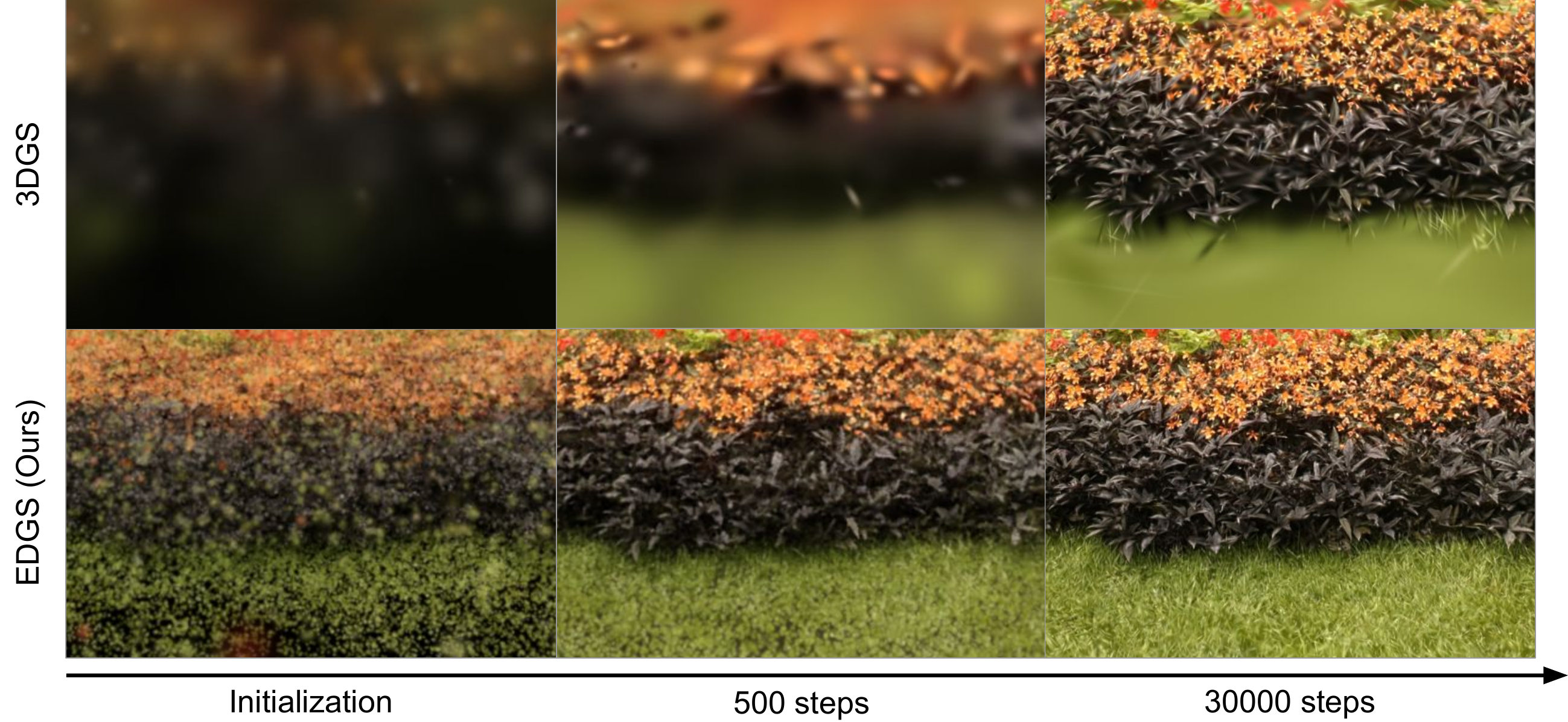} 
       \caption{EDGS accelerates Gaussian Splatting by replacing incremental densification with a dense initialization of splats inferred from 2D correspondences. This leads to faster convergence and higher rendering quality. \textit{Left:} on MipNeRF360~\cite{barron_mip-nerf_2022}, EDGS reaches the original 3DGS~\cite{kerbl20233d} quality in 15\% of training time and achieves 35\% lower LPIPS when trained further, outperforming recent acceleration and quality-focused methods~\cite{mallick2024taming, kheradmand20243d}. Reported time includes initialization. \textit{Right:} renderings closely match ground truth after only 500 steps. }
    \label{fig:teaser_arxiv}
    \end{center}
    }]
\def\thefootnote{*}\footnotetext[0]{Equal contribution}
\begin{abstract}
3D Gaussian Splatting reconstructs scenes by starting from a sparse Structure‑from‑Motion initialization and refining under‑reconstructed regions. This process is slow, as it requires multiple densification steps where Gaussians are repeatedly split and adjusted, following a lengthy optimization path. Moreover, this incremental approach often yields suboptimal renderings in high-frequency regions.

We propose a fundamentally different approach: eliminate densification with a one-step approximation of scene geometry using triangulated pixels from dense image correspondences. This dense initialization allows us to estimate the rough geometry of the scene while preserving rich details from input RGB images, providing each Gaussian with well-informed color, scale, and position. As a result, we dramatically shorten the optimization path and remove the need for densification. Unlike methods that rely on sparse keypoints, our dense initialization ensures uniform detail across the scene, even in high-frequency regions where other methods struggle. Moreover, since all splats are initialized in parallel at the start of optimization, we remove the need to wait for densification to adjust new Gaussians.

EDGS reaches LPIPS and SSIM performance of standard 3DGS significantly faster than existing efficiency-focused approaches. When trained further, it exceeds the reconstruction quality of state-of-the-art models aimed at maximizing fidelity. Our method is fully compatible with other acceleration techniques, making it a versatile and efficient solution that can be integrated with existing approaches. 
\end{abstract}    
\section{Introduction}
Reconstructing 3D scenes from collections of 2D images is a fundamental challenge in computer vision~\cite{hartley2003multiple, 6619014, mildenhall2021nerf}, with applications in virtual and augmented reality~\cite{Janai2017ComputerVF, LI2022102298, saito2024relightable, qi2017pointnet, zielonka2023drivable, guedon2024sugar},  robotics~\cite{ozyesil2017surveystructuremotion, yan2024gs, lu2024maniGaussian}, and content creation~\cite{article, Baatz2022NeRFTexNR, Chen2023TeSTNeRFT3, wast}. The goal is to obtain high-quality 3D representations efficiently, enabling real-time rendering while maintaining reconstruction fidelity. However, achieving balance between efficiency, speed, and quality requires a representation that is both expressive and computationally efficient. NeRF-based models~\cite{mildenhall2021nerf, yu2021plenoctrees, barron_mip-nerf_2022, muller2022instant, yu2021pixelnerf, garbin2021fastnerf} control the trade-off between quality, computational cost, and representation capacity by designing network architectures and increasing the number of parameters. In contrast, point-based graphics~\cite{hedman2018deep, yariv_bakedsdf_2023, qi2017pointnet} represent surfaces using discrete primitives, such as meshes or point clouds, offering more direct control over complexity but often struggling with quality and scalability.

Recently, 3D Gaussian Splatting (3DGS)~\cite{kerbl20233d} has emerged as a powerful and efficient alternative for representing 3D scenes. It models scenes as a set of optimized 3D Gaussians, mathematical primitives defined by their position, color, and spread. The method begins with sparse initialization, typically derived from Structure-from-Motion~(SfM)~\cite{schonberger2016structure}, and progressively refines scene by adding splats to under-reconstructed regions. Through this densification process, 3DGS reaches high rendering quality while efficiently allocating computational resources.

\begin{figure}
    \centering
    \captionsetup{type=figure}
        \includegraphics[width=0.99\linewidth]{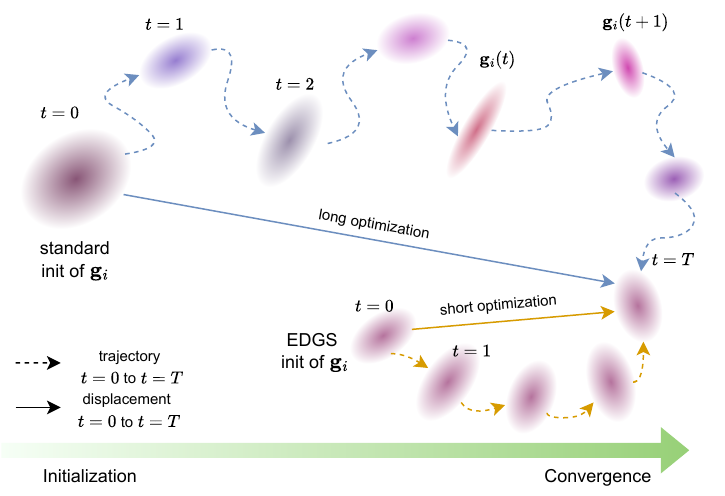}
  \caption{
 EDGS initializes Gaussians closer to their final positions, resulting in shorter optimization trajectories and faster convergence to high-quality reconstructions.
 }
  \label{fig:colorcoord_motiv}
\end{figure}

However, this process is suboptimal. The original 3DGS detects under-reconstructed regions using the gradient norm of the photometric loss. But this metric often fails in high-frequency regions and does not align well with human perception. A separate branch of papers has proposed pixel-error-driven formulations~\cite{bulo2024revising, zhang2024pixel, cheng2024Gaussianpro, mallick2024taming}, gradient calculation improvements~\cite{ye2024absgs}, and even treating 3DGS as Markov Chain Monte Carlo samples~\cite{kheradmand20243d}. Despite these efforts, accurately capturing fine details, particularly in high-frequency regions, remains a challenge, as illustrated in~\cref{fig:teaser_arxiv}. Furthermore,  while each densification step is computationally efficient, the overall process is slow. It requires many update steps, as Gaussians must iteratively adjust their parameters before the model determines that additional splats are necessary. This results in a long optimization path, where individual Gaussians undergo multiple refinements before reaching their final states (see~\cref{sec:abl},~\cref{fig:colorcoord_motiv} and~\cref{fig:colorcoor_adjustment}). Densification delays convergence, as it takes many iterations for the model to identify areas requiring higher reconstruction fidelity. These challenges raise an important question: \textit{can we bypass densification entirely?}

In this paper, we propose a direct initialization strategy that eliminates the need for incremental densification used in the original 3DGS method. Rather than waiting for the model to gradually fill in missing details, we precompute a dense set of 3D Gaussians by triangulating dense 2D correspondences across multiple input views. Knowing the viewing rays for each correspondence pixel and the camera poses, we recover 3D positions of Gaussians by triangulating matched pixels between image pairs. This allows us to assign each Gaussian well-informed initial parameters, such as position, color, and scale, from the start. To summarize, we replace the slow iterative densification process of the scene with a dense initialization. As a result, each Gaussian is immediately supervised by rich per-pixel photometric signal, allowing for efficient optimization of the entire scene.

Although this initialization is noisy (see~\cref{fig:teaser_arxiv}), we show that it remains robust and leads to faster convergence. Our experiments, quantitatively and qualitatively, confirm that EDGS yields higher reconstruction quality, shorter training time, fewer Gaussians, and eliminates the need for densification. Our contributions can be summarized as follows:
\begin{itemize}
    \item We introduce a novel dense initialization for 3D Gaussian Splatting, based on the sampling distribution of triangulated multi-view correspondences, which effectively replaces the traditional incremental refinement process.
    \item EDGS achieves faster convergence and higher reconstruction quality than prior 3DGS methods. We further analyze how the proposed initialization affects the optimization trajectories of individual Gaussians.
    \item Our initialization improves reconstruction without modifying the optimization algorithm, making it compatible with other 3DGS methods. This makes it a complementary component that can be seamlessly integrated with adaptive densification strategies to further enhance performance.
\end{itemize}  
\section{Related Work}

\noindent\textbf{Novel View Synthesis} generates images from new viewpoints. A breakthrough in this area was \mbox{Neural} Radiance \mbox{Fields (NeRF)~\cite{mildenhall2021nerf}}, which reconstructs 3D scenes from 2D images using volumetric rendering techniques~\cite{drebin1988volume, levoy1990efficient, max1995optical, max2005local}. Follow‑ups have adapted NeRF to sparse views~\cite{irshad2023neo, yu2021pixelnerf, wang2023sparsenerf, kim2022infonerf, niemeyer2022regnerf}, sped up rendering~\cite{garbin2021fastnerf, sun2022direct, liu2020neural, yu2021plenoctrees}, and cut training time~\cite{reiser2021kilonerf, muller2022instant, niemeyer2022regnerf, xu2022sinnerf}. Despite these gains, sampling points along a ray and passing them through an MLP introduces slowdowns. In contrast, 3D Gaussian Splatting (3DGS)~\cite{kerbl20233d} delivers an explicit representation with high fidelity and real‑time performance. It has proven effectiveness for human avatars~\cite{kocabas2024hugs, lei2024gart, saito2024relightable, zielonka2023drivable}, text‑to‑3D generation~\cite{chen2024text, tang2023dreamGaussian, yi2023Gaussiandreamer}, dynamic scenes modeling~\cite{luiten2023dynamic, wu20244d, yang2023real, duan20244d, katsumata2023efficient, yang2024deformable}, and more~\cite{wang2021neus, wang2023neus2, xiao2023level, yariv2021volume, guedon2024sugar, xie2024physGaussian, ye2023Gaussian}. However, it still struggles with aliasing~\cite{yu2024mip, yan2024multi}, memory usage~\cite{niedermayr2024compressed, navaneet2023compact3d, lee2024compact, lu2024scaffold, girish2023eagles}, surface reconstruction~\cite{guedon2024sugar, huang20242d}, and complex regions~\cite{yang2024spec}.

\paragraph{Densification} Several studies suggest that using an effective strategy for splat densification can significantly enhance performance. RevDev~\cite{bulo2024revising} introduced a per-pixel error function as a criterion for densification. \mbox{AbsGS~\cite{ye2024absgs}} addressed the issue of gradient collision during the detection of under-reconstructed regions. \mbox{MiniSplatting~\cite{Fang2024MiniSplattingRS}} proposed a densification approach that incorporates both screen-space and world-space information. ScaffoldGS~\cite{lu2024scaffold} introduced anchor points and a growth algorithm to optimize Gaussians distribution. Meanwhile, 3DGS-MCMC~\cite{kheradmand20243d} reformulated densification as a Markov Chain Monte Carlo sampling process, enabling an efficient Gaussian distribution across scene.
In contrast, we propose an improved initialization method that avoids densification altogether, eliminating the need to detect under-reconstructed regions.

\paragraph{Efficiency} A number of recent works aim to enhance the efficiency of 3DGS. One approach leverages pre-trained neural networks as priors to guide reconstruction~\cite{chen2025mvsplat, zou2024triplane, fan2024instantsplat, xu2024depthsplat}. This data-driven strategy enables rapid reconstruction with high quality, particularly in sparse-view scenarios. We focus on dense-view reconstruction. Another area of research targets the optimization of 3DGS by refining the differentiable rasterizer~\cite{durvasula2023distwar, mallick2024taming, feng2024flashgs} or improving the framework itself~\cite{ye2024gsplatopensourcelibraryGaussian}. Separately, 3DGS-LM~\cite{hollein20243dgs} proposes a Levenberg-Marquardt optimizer that integrates with the 3DGS rasterizer and can be adapted to other rasterization methods.  Our approach instead focuses on improving the initialization process, which is compatible with these optimizations.

\paragraph{Initialization} Recent works, such as RAIN-GS~\cite{jung2024relaxing} and 3DGS-MCMC~\cite{kheradmand20243d}, shows that random initialization can match the performance of 3DGS. In contrast, RadSplat~\cite{niemeyer2024radsplat} initializes from points extracted using pretrained NeRFs to improve quality, though it requires 9 hours of training. EDGS departs from both approaches by emphasizing efficiency while outperforming quality-focused methods.
\section{Method}

\label{sec:intuition}
Our goal is to initialize a dense set of Gaussian splats~(\cref{sec:prelim}).
Instead of incrementally adding 
information via photometric loss, we leverage all available 2D image information from the start~(\cref{sec:dense_matching}).
We first triangulate dense pixel correspondences into 3D space~(\cref{sec:splat_init}).
Then, we aggregate semantic confidence and geometric consistency across neighboring views to build a sampling distribution $\mathbf{p}^i$ for each reference image~(\cref{sec:weak_to_dense}), from which splats are sampled.
Finally, we initialize spherical harmonics for the sampled splats~(\cref{sec:sh}).
\subsection{Preliminaries}
\label{sec:prelim}
3DGS~\cite{kerbl20233d} represents scenes as collections of Gaussians $\mathbb{G} = \bigcup_{i=1}^{N} \bm{g}_i$, rendered into images using a splatting-based rasterization technique~\cite{zwicker2002ewa}. Each Gaussian component $ \bm{g}_i $ is described by parameters $\{ \bm{g}^x_i, \bm{\Sigma}_i, \bm{g}^c_i, \bm{g}^{\alpha}_i \}$ for $ i \in \{1, \ldots, N\} $. Specifically, $\bm{g}^x_i \in \mathbb{R}^3$ is the center of the Gaussian $\bm{g}_i$ in 3D space, $\bm{\Sigma}_i \in \mathbb{R}^7$ encodes its shape, $\bm{g}^c_i \in \mathbb{R}^3$ defines its RGB color, and $\bm{g}^{\alpha}_i \in \mathbb{R}^1$ indicates its opacity. The color $C$ of a given pixel $p$ is rendered as:
\begin{equation}
    \begin{aligned}
    \label{eq:prelim}
        C(p) = \sum_{i=1}^{N} \bm{g}^{c}_i \bm{\sigma}_i (p) \prod_{j=1}^{i-1} (1 - \bm{g}^{\alpha}_j); \\
        \bm{\sigma}_i(p) = \bm{g}^{\alpha}_i e^{-\frac{1}{2} (\bm{p}' - \bm{g}^{x}_i)^T \bm{\Sigma}_i^{-1} (\bm{p}' - \bm{g}^{x}_i)},
    \end{aligned}
\end{equation}
where $\bm{\sigma}_i$ measures the influence of the $i$-th Gaussian on pixel $p$, with $(\bm{p}' - \bm{g}^x_i)$ representing the shortest distance between the pixel projection line and the Gaussian center $\bm{g}^x_i$.

To project 3D Gaussians to 2D for rendering, following~\cite{kerbl20233d}, we reparameterize the covariance matrix $\bm{\Sigma}_i$ as a function of scaling~$\bm{S}_i$ and rotation~$\bm{R}_i$ matrices ensuring the positive semi-definiteness of $\bm{\Sigma}_i$:
\begin{equation}
 \label{eq:prelim_scaling}
    \bm{\Sigma}_i = \bm{R}_i \bm{S}_i \bm{S}_i^T \bm{R}_i^T.
\end{equation}
Gaussians $\mathbb{G}$ are optimized with photometric loss.

\subsection{Extract information from 2D prior}
\label{sec:dense_matching}
We begin by selecting a reference image $I^i$ from the training set. For each $I^i$, we identify neighboring images $\mathbb{I}_i = \{I^1, \dots, I^j | j \in [0, J]\}$ that have maximal overlap with $I^i$, based on camera parameters and spatial proximity. We measure proximity between camera matrices using the Frobenius norm.

For each neighboring image $I^j \in \mathbb{I}_i$, we compute dense correspondences in $I^i$ using a pretrained network $\mathcal{M}$. This network estimates pixel-wise correspondences between $I^i$ and $I^j$ as:
\begin{equation}\label{eq:methods:roma:goal}
\mathcal{M}(I^i, I^j) \to (\mathcal{W}^{i \rightarrow j}, \mathbf{c}^{ij}),
\end{equation}
where $\mathcal{W}^{i \rightarrow j} \in \mathbb{R}^{2 \times H \times W}$ is a dense forward warp field mapping pixels from $I^i$ to $I^j$, and $\mathbf{c}^{ij} \in \mathbb{R}^{H \times W}$ encodes correspondence confidence. For a pixel $(u^i_k, v^i_k) \in I^i$ with index $k$ , the warp provides its mapped location in $I^j$ via $\mathcal{W}^{i \rightarrow j}(u^i_k, v^i_k)$.

\subsection{Splats triangulation}
\label{sec:splat_init}
 The goal is to find an accurate 3D position of a Gaussian splat $\bm{g}_k^x = (x_k, y_k, z_k) \new{ \in \mathbb{R}^{1\times 3}}$ corresponding to matched keypoint pair $(u_k^i, v_k^i)$ and $(u_k^j, v_k^j)$. 

We are given projection matrices $\bm{P}^i, \bm{P}^j \in \mathbb{R}^{4\times3}$ for cameras $i,j$, which map 3D homogeneous coordinates to 2D homogeneous coordinates: 

\begin{equation}
\begin{aligned}
\begin{cases}
\left[ \bm{g}_k^x \;\; 1 \right] \bm{P}^i = w_k^i \left[ u_k^i \;\; v_k^i \;\; 1 \right], \\
\left[ \bm{g}_k^x \;\; 1 \right] \bm{P}^j = w_k^j \left[ u_k^j \;\; v_k^j \;\; 1 \right],
\end{cases}
\end{aligned}  
\end{equation}
where $w^i_k$ and $w^j_k$ are scalars for homogeneous coordinate normalization.
From the projection equations for camera $i$ we obtain: $\left[ \bm{g}_k^x, 1 \right]^T \bm{P}^i_{\text{col},0} = w_k^i u_k^i$, $\left[ 
\bm{g}_k^x, 1 \right]^T \bm{P}^i_{\text{col},1} = w_k^i v_k^i$. But we also know that third column gives us $\left[ \bm{g}_k^x, 1 \right]^T \bm{P}^i_{\text{col},2} = w_k^i$. Substituting the last expression for $w_k^i$ into the first two yields:
\begin{equation}
\begin{aligned}
\begin{cases}
\left[ \bm{g}_k^x \;\; 1 \right] \bm{P}^i_{\text{col},0} - u_k^i \left[ \bm{g}_k^x \;\; 1 \right] \bm{P}^i_{\text{col},2} = 0, \\
\left[ \bm{g}_k^x \;\; 1 \right] \bm{P}^i_{\text{col},1} - v_k^i \left[ \bm{g}_k^x \;\; 1 \right] \bm{P}^i_{\text{col},2} = 0, \\
\left[ \bm{g}_k^x \;\; 1 \right] \bm{P}^j_{\text{col},0} - u_k^j \left[ \bm{g}_k^x \;\; 1 \right] \bm{P}^j_{\text{col},2} = 0, \\
\left[ \bm{g}_k^x \;\; 1 \right] \bm{P}^j_{\text{col},1} - v_k^j \left[ \bm{g}_k^x \;\; 1 \right] \bm{P}^j_{\text{col},2} = 0.
\end{cases}
\end{aligned}
\end{equation}

We rearrange the equations to the form \mbox{$A \bm{g}_k^x = -b$}, where $A$ is constructed from the projection matrices and  $b$ being a vector of constants:
\begin{equation}
\begin{aligned}
  A^T = \begin{bmatrix}
        \bm{P}^i_{\text{col},0} - u_k^i \bm{P}^i_{\text{col},2} \\
        \bm{P}^i_{\text{col},1} - v_k^i \bm{P}^i_{\text{col},2} \\
        \bm{P}^j_{\text{col},0} - u_k^j \bm{P}^j_{\text{col},2} \\
        \bm{P}^j_{\text{col},1} - v_k^j \bm{P}^j_{\text{col},2}
    \end{bmatrix}, \quad
    b = \begin{bmatrix} 0 \\ 0 \\ 0 \\ 0 \end{bmatrix}.
\end{aligned}
\end{equation}
\begin{equation}
\begin{aligned}
    \bm{g}_k^x := \arg\min_{\bm{x}} \|A \bm{x} + b\|^2
\end{aligned}
\end{equation}
to obtain a solution of coordinates for Gaussian $k$ in homogeneous coordinates $\bm{g}_k^x = [x_k, y_k, z_k, 1]^T$.

\subsection{Sampling distribution}
\label{sec:weak_to_dense}
Given triangulated correspondences between a reference view $I^i$ and a neighboring view $I^j$, directly using all matches for multi-view reconstruction is computationally infeasible. We therefore define a sampling distribution $\mathbf{p}^i$ to select geometrically consistent and semantically reliable correspondences. For each triangulated 3D point $\bm{g}^x_k$, we compute its reprojection error in the reference image $I^i$:
\begin{equation}
\varepsilon_k^i =
\bigl\|\pi(\bm{P}^i,\bm{g}^x_k) - (u^i_k,v^i_k)\bigr\|_2
\end{equation}
where $\pi(\bm{P},\cdot)$ denotes projection with camera matrix $\bm{P}$; $\varepsilon_k^j$ is defined analogously. Points with high reprojection $\varepsilon_k^{ij}:=\text{max}(\varepsilon_k^i,\varepsilon_k^j)$ error are likely inconsistent across views and should be avoided. We convert reprojection errors  and correspondence confidences $\mathbf{c}^{ij}(u^i_k,v^i_k)$ over some threshold $\tau_{\text{corr}}$ into uniform sampling distributions:
\begin{equation}
\mathbf{p}^{ij}_{\text{corr}} \sim \mathcal{U}\!\left(\{k \mid \mathbf{c}^{ij}(u^i_k,v^i_k) > \tau_{\text{corr}}\}\right),
\end{equation}
\begin{equation}
\mathbf{p}^{ij}_{\text{proj}} \sim \mathcal{U}\!\left(\{k \mid \varepsilon^{ij}_k < \tau_{\text{proj}}\}\right).
\end{equation}
Finally, we combine the geometry-based and confidence-based probabilities via element-wise multiplication for all nearest neighbors $\mathbb{I}_i$:
\begin{equation}\label{eq:methods:sampling}
\mathbf{p}^i(k)
\;\propto\; \max_{j \in \mathbb{I}_i}
\left(
\mathbf{p}_{\text{corr}}^{ij}(k)
\mathbf{p}_{\text{proj}}^{ij}(k)
\right),
\end{equation}
which prioritizes points with the strongest geometric and correspondence consistency across neighbors.

We then form the global sampling distribution by aggregating the per-reference $I^i$ probabilities
$\mathbf{p}(k) \propto \bigcup_i \mathbf{p}^i_k$,
effectively selecting correspondences that remain consistent across multiple reference views. 

\subsection{Spherical harmonics}
\label{sec:sh}

After sampling Gaussians from $\mathbf{p}^i$, we assign each splat an initial color from the reference image $I^i$ at pixel coordinates $(u_k^i, v_k^i)$. For each splat, we collect $n$ RGB observations $\mathbf{O}_k \in \mathbb{R}^{n \times 3}$ from view directions $\mathbf{v}_1, \dots, \mathbf{v}_n \in \mathbb{R}^3$ and estimate the rest of its spherical harmonics (SH) coefficients. We build a matrix $\mathbf{Y}_k \in \mathbb{R}^{n \times 16}$, where each row contains the 16 real SH basis functions (up to degree $l=3$) evaluated at direction $\mathbf{v}_i$. The SH coefficients $\hat{\mathbf{H}}_k \in \mathbb{R}^{16 \times 3}$ are obtained by solving the system of linear equations: 
\begin{equation}
    \hat{\mathbf{H}}_k = \arg\min_{\mathbf{H} \in \mathbb{R}^{16 \times 3}} \bigl\| \mathbf{Y}_k \mathbf{H} - \mathbf{O}_k \bigr\|_F^2.
\end{equation}

When $n < 16$, we use the Moore–Penrose pseudoinverse~\cite{Penrose_1955},
\begin{equation}
\hat{\mathbf{H}}_k = \mathbf{Y}_k^{+}\,\mathbf{O}_k,
\end{equation}
ensuring stable estimation under limited observations.

Finally, these initialized Gaussians undergo standard photometric loss optimization to refine their parameters and achieve precise 3D reconstructions.

\section{Experiments}

\begin{table*}[t]
\setlength{\abovecaptionskip}{6pt}
\setlength{\belowcaptionskip}{0pt}
    \centering  
    \rowcolors{4}{gray!10}{white}
    \adjustbox{max width=\textwidth}{\begin{tabular}{llcc @{\hskip 2mm} c @{\hskip 2mm} c l l | c@{\hskip 2mm} c @{\hskip 2mm}c l l | c @{\hskip 2mm} c @{\hskip 2mm} c  l l}
        \toprule
       & & \multirow{2}{*}[-7pt]{%
  \begin{tabular}[c]{@{}c@{}}%
    {\small \textbf{Densifica-}}\\
    {\small \textbf{tion free}} \\
  \end{tabular}%
}  &\multicolumn{5}{c}{\textbf{Tanks $\&$ Temples}} & \multicolumn{5}{c}{\textbf{Mip-NeRF 360}} &  \multicolumn{5}{c}{\textbf{Deep Blending}}  \\
        \cmidrule(lr){4-18} 
                       & &  & \textbf{SSIM} $\uparrow$ & \textbf{PSNR} $\uparrow$ & \textbf{LPIPS} $\downarrow$ & \multicolumn{1}{c}{\begin{tabular}[c]{@{}c@{}}\textbf{Train}\\ \textbf{time}\end{tabular}} &  \multicolumn{1}{c|}{\begin{tabular}[c]{@{}c@{}}\textbf{\#G} \\ ($10^6$)\end{tabular}} & 
                          \textbf{SSIM} $\uparrow$& \textbf{PSNR}$\uparrow$ & \textbf{LPIPS}$\downarrow$ & \multicolumn{1}{c}{\begin{tabular}[c]{@{}c@{}}\textbf{Train}\\ \textbf{time}\end{tabular}} &  \multicolumn{1}{c|}{\begin{tabular}[l]{@{}l@{}}\textbf{\#G} \\ ($10^6$)\end{tabular}} &  
                          \textbf{SSIM}$\uparrow$ & \textbf{PSNR} $\uparrow$& \textbf{LPIPS} $\downarrow$ & \multicolumn{1}{c}{\begin{tabular}[c]{@{}c@{}}\textbf{Train}\\ \textbf{time}\end{tabular}} &  \multicolumn{1}{l}{\begin{tabular}[c]{@{}c@{}}\textbf{\#G} \\ ($10^6$)\end{tabular}} \\
        \midrule
         \cellcolor{white}& Plenoxels~\cite{fridovich-keil_plenoxels_2022}  &  {\color{ourgreen}\cmark}  & 0.719  & 21.08  & 0.379  & 25 m  & \ \ -  & 0.626  & 23.08  & 0.463  & 26 m  & \ \ -   & 0.795  & 23.06  & 0.510  & 28 m  &  \ \ -  \\
         \cellcolor{white}& INGP-Big~\cite{muller2022instant} &   {\color{ourgreen}\cmark}  & 0.745  & 21.92  & 0.305  & \textbf{7 m } & \ \ - & 0.699  & 25.59  & 0.331  & \textbf{8 m }  & \ \ -   & 0.817  & 24.96  & 0.390  & \textbf{ 8 m}  & \ \ -  \\
          \cellcolor{white} \multirow{-3}{*}{\rotatebox{90}{\textbf{Rays}}}  & Mip-NeRF360~\cite{barron_mip-nerf_2022} &  {\color{ourgreen}\cmark} & 0.759  & 22.22  & 0.257  & 48 h  & \ \ - & 0.792  & 27.69  & 0.237  & 48 h    & \ \ -   & 0.901  & 29.40  & 0.245  & 48 h  & \ \ -  \\
  
        \midrule
        \cellcolor{white}&3DGS~\cite{kerbl20233d} &  {\color{ourred}\xmark}   & 0.841  & 23.14  & 0.183  & 27 m$^{\ast\ast}$  & 2.0       & 0.815  & 27.21  & 0.214  & 42 m$^{\ast\ast}$    & 3.5  & 0.903  & 29.41  & 0.243  & 36 m$^{\ast\ast}$  & 3.2  \\
        \cellcolor{white}&3DGS~\cite{kerbl20233d}* &  {\color{ourred}\xmark}  & 0.853  & 23.76 & 0.169  & 19 m  & 1.6     & 0.816  & 27.49  & 0.215  & 26 m    & \underline{2.8}  & \textbf{0.908}  & 29.77  & 0.242  &  27 m &  2.6 \\
        \cellcolor{white}& AbsGS-0004~\cite{ye2024absgs} &  {\color{ourred}\xmark} & 0.852  & 23.59  & 0.162  & 14 m  & \textbf{1.4}   & 0.818  & 27.41  & 0.198 & \underline{20 m}    & 3.1    & 0.901  & 29.61  & \underline{0.236}  & 20 m  & \underline{1.9}  \\
        \cellcolor{white} & Rain-GS~\cite{jung2024relaxing}$^{\dagger}$ &  {\color{ourred}\xmark} &  0.823 & 23.13 & 0.207  & 15 m$^{\star \star}$  & \ \ -  &  0.807  & 27.23  & 0.229  &  32 m$^{\star \star}$    & \ \ -   & 0.900  &  29.42 & 0.255  & 28 m$^{\star \star}$  & \ \ - \\
        \cellcolor{white} & Mip-Splatting~\cite{yu2024mip} &  {\color{ourred}\xmark} & 0.859  & 23.81  & \underline{0.156}  & 16 m  & 2.4  & 0.838  & 27.97  & 0.179  &  26 m   & 4.0  &  0.903 & 29.35 & 0.239  & 29 m  &  3.6 \\
        \cellcolor{white} & 3DGS-MCMC~\cite{kheradmand20243d} &  {\color{ourred}\xmark} &  \underline{0.863} & \underline{24.22}  & 0.158  & \underline{13 m}  & \underline{1.9}  &  \textbf{0.842} &  \textbf{28.15} &   \underline{0.176}   &  \underline{20 m} & 3.2   & 0.902  & 29.56  & 0.244  & \underline{19 m} & 2.9 \\
        \cellcolor{white}& ScaffoldGS~\cite{lu2024scaffold} &  {\color{ourred}\xmark}  & 0.854 & 24.08  & 0.165  & 23 m  & 6.0$^\ddagger$   & 0.812  & 27.60  & 0.222  & 22 m    & 6.0$^\ddagger$  & \underline{0.907}  &  \textbf{30.25} & 0.245  & 28 m  & 4.0$^\ddagger$  \\

        \cmidrule(lr){2-18}
        \cellcolor{white} \multirow{-7}{*}{\rotatebox{90}{\textbf{ \quad Gaussians}}}&\cellcolor{white}  \textbf{EDGS + 3DGS} &  \cellcolor{white} {\color{ourgreen}\cmark} & \cellcolor{white} \textbf{0.868} & \cellcolor{white} \textbf{24.28} &  \cellcolor{white} \textbf{0.132} & \cellcolor{white} 23 m & \cellcolor{white} \textbf{1.4} & \cellcolor{white} \underline{0.839} & \cellcolor{white} \underline{28.02} & \cellcolor{white} \textbf{0.141} & \cellcolor{white} 27 m & \cellcolor{white} \textbf{1.9}   &
                         \cellcolor{white} 0.904  & \cellcolor{white} \underline{29.81} & \cellcolor{white} \textbf{0.223} & \cellcolor{white} 30 m & \cellcolor{white} \textbf{1.6} \\    

        \bottomrule
       \multicolumn{18}{c}{\cellcolor{white} {\small $^\dagger$ results were taken directly from the paper, $^\ddagger$ for ScaffoldGS denotes the number of splats produced by anchors, $^{\ast \ast}$ results reported for an NVIDIA A6000, $^{\star \star}$ results or an NVIDIA RTX 3090.}} \\
    \end{tabular}}

   \caption{Performance comparison on Mip-NeRF 360~\cite{barron_mip-nerf_2022}, Tanks$\&$Temples~\cite{knapitsch2017tanks}, and Deep Blending~\cite{hedman2018deep}. Our method \textit{improves reconstruction quality} across all benchmarks while using the standard 3DGS optimization pipeline \textit{without densification}. Reported training times for our approach include both initialization and optimization, whereas for other methods (except Rain-GS~\cite{jung2024relaxing}) only the optimization time is counted. Checkmarks in the \textit{Densif. free} column indicate models not using densification. Detailed per-scene results are provided in~\cref{sec:sup:per_scene_results}.}
    \label{tab:merged_comparison}
\end{table*}

\begin{table}[t]
    \centering  
    \adjustbox{max width=0.47\textwidth}{\begin{tabular}{c c c @{\hskip 2mm} c @{\hskip 2mm} c @{\hskip 2mm} c @{\hskip 2mm}c l }
    \toprule
     \multicolumn{2}{c}{\cellcolor{white}\textbf{Mip-NeRF 360}} & \multicolumn{1}{c}{\begin{tabular}[c]{@{}c@{}}\textbf{Densif.}\\ \textbf{free}\end{tabular}}& \textbf{SSIM} $\uparrow$ & \textbf{PSNR} $\uparrow$ & \textbf{LPIPS} $\downarrow$ & \multicolumn{1}{c}{\begin{tabular}[c]{@{}c@{}}\textbf{Train}\\ \textbf{time}\end{tabular}} &  \multicolumn{1}{c}{\begin{tabular}[c]{@{}c@{}}\textbf{\#G} \\ ($10^6$)\end{tabular}}  \\
    \midrule
    & gsplat~\cite{ye2024gsplatopensourcelibraryGaussian} &  {\color{ourred}\xmark}   & 0.818  & 27.51  & 0.215  & 18 m    & 3.1  \\
   \cellcolor{white} & 3DGS+3DGS-LM~\cite{hollein20243dgs}$^\dagger$ &  {\color{ourred}\xmark} &   0.813  & 27.39  & 0.221  & 16 m    & 2.8$^{\star}$ \\
    & EAGLES~\cite{girish2023eagles} &  {\color{ourred}\xmark}  & 0.809 &  27.20 & 0.232  &  16 m   &  \underline{1.3}   \\
    \cellcolor{white} & Taming 3DGS~\cite{mallick2024taming} &  {\color{ourred}\xmark}  & 0.820  & \textbf{27.71}  & 0.207  &  14 m   & 3.2   \\
    & MiniSplatting~\cite{Fang2024MiniSplattingRS} &  {\color{ourred}\xmark}  & 0.820  &  27.25 & 0.217  &  \underline{12 m}   & \textbf{0.5}  \\
    & \textbf{EDGS + 3DGS 10K} & {\color{ourgreen}\cmark} & \textbf{0.834} & \underline{27.54} & \textbf{0.154} & \underline{12 m} & 2.1 \\
   \multirow{-8}{*}{
\cellcolor{white}
\makebox[0pt][c]{%
\begin{tikzpicture}[baseline=(A.base)]
  \node (A) {};
  \draw[->, ultra thick] (A.south) -- ++(0,-2.5)
    node[midway, above=7pt, font=\bfseries, rotate=90,]{Faster};
\end{tikzpicture}
}}  & \textbf{EDGS + 3DGS 5K}  & {\color{ourgreen}\cmark} & \underline{0.825} & 26.88 & \underline{0.166} & \textbf{8 m} & 2.6
                      \\

        \bottomrule
         \multicolumn{8}{c}{\cellcolor{white} {\small 
         $^\dagger$~From original paper. $^\star$~Assumed same as 3DGS~\cite{kerbl20233d} due to identical densification.}} 
    \end{tabular}}
    \caption{Quantitative evaluation under early-stopping settings on Mip-NeRF 360~\cite{barron_mip-nerf_2022}. When trained for a comparable duration (\textit{EDGS 10K}), our method outperforms other efficient approaches. Even with reduced training time (\textit{EDGS 5K}), it outperforms all competing methods on two of the three standard metrics.}
    \label{tab:efficciency_comparison}
\end{table}

\begin{table}[ht]
    \centering  
    \adjustbox{max width=0.47\textwidth}{\begin{tabular}{l@{\hskip 2mm} c @{\hskip 2mm} l @{\hskip 2mm} l@{\hskip 2mm} l @{\hskip 2mm} l  l}
        \toprule
                     \textbf{Mip-NeRF 360} & \multicolumn{1}{c}{\begin{tabular}[c]{@{}c@{}}\textbf{EDGS}\\ \textbf{Init}\end{tabular}} & \textbf{SSIM} $\uparrow$ & \textbf{PSNR} $\uparrow$ & \multicolumn{1}{c}{\textbf{LPIPS} $\downarrow$} & \multicolumn{1}{c}{\begin{tabular}[c]{@{}c@{}}\textbf{Train}\\ \textbf{time}\end{tabular}} &  \multicolumn{1}{c}{\begin{tabular}[c]{@{}c@{}}\textbf{\#G} \\ ($10^6$)\end{tabular}} \\
        \midrule
         &  {\color{ourred}\xmark}  & 0.818  & 27.41  & 0.198 & 20 m    & 3.1 \\
         \multirow{-2}[0]{*}{AbsGS-0004~\cite{ye2024absgs}} &  {\color{ourgreen}\cmark}  & 0.822\textsubscript{\color{ourgreen}$\blacktriangle1\%$} & 27.53\textsubscript{\color{ourgreen}$\blacktriangle 0.12$} & 0.187\textsubscript{\color{ourgreen}$\blacktriangledown$6\%} & 19 m & 3.0  \\
        \rowcolor{gray!10}  &  {\color{ourred}\xmark} &  0.842 &  28.15 &   0.176   &  20 m & 3.2 \\
        \rowcolor{gray!10} \multirow{-2}[0]{*}{3DGS-MCMC~\cite{kheradmand20243d}} &  {\color{ourgreen}\cmark} & 0.847\textsubscript{\color{ourgreen}$\blacktriangle 1\%$} & 28.29\textsubscript{\color{ourgreen}$\blacktriangle 0.14$} & 0.159\textsubscript{\color{ourgreen}$\blacktriangledown$10\%} & 20 m & 3.2  \\
        &  {\color{ourred}\xmark} & 0.820  & 27.71  & 0.207  &  14 m   & 3.2 \\
       \multirow{-2}[0]{*}{Taming 3DGS~\cite{mallick2024taming}} &  {\color{ourgreen}\cmark}  & 0.842\textsubscript{\color{ourgreen}$\blacktriangle 3\%$} & 28.07\textsubscript{\color{ourgreen}$\blacktriangle 0.36$} & 0.179\textsubscript{\color{ourgreen}$\blacktriangledown$14\%} & 11 m & 3.2 \\
        \bottomrule
    \end{tabular}}

    \caption{EDGS as initialization for different densification methods. Incorporating our approach, without fine-tuning hyperparameters and across various settings, consistently improves all other methods without increasing final Gaussian count or increasing training time. For \textit{EDGS Init}, the reported time includes the initialization phase. }
    \label{tab:ours_plus_adc}
\end{table}

\subsection{Datasets and Metrics}
We evaluate on Mip-NeRF360~\cite{barron_mip-nerf_2022}, Tanks\&Temples~\cite{knapitsch2017tanks}, and Deep Blending~\cite{hedman2018deep} datasets. Following standard protocol, we use 9, 2, and 2 scenes, respectively. Evaluation metrics include PSNR, SSIM~\cite{ssim}, and LPIPS~\cite{zhang2018unreasonable}, along with training time and final Gaussians count. All experiments were run on NVIDIA A100 GPUs, with competing methods re-evaluated on the same hardware for fairness. Runtimes for EDGS include initialization (see~\cref{sec:sup:implement_det} for time break-down).

\subsection{Baselines}
\label{sec:baselines}
For ray‑based methods, we include Plenoxels~\cite{yu2021plenoctrees}, Mip‑NeRF360~\cite{barron_mip-nerf_2022}, and Instant‑NGP~\cite{muller2022instant}. As our method is based on 3DGS, we also compare with the original 3DGS~\cite{kerbl20233d}. We retrain it (denoted as 3DGS*), as this resulted in better performance than the originally reported scores. We also include high-quality baselines \mbox{AbsGS}\cite{ye2024absgs}, Mip-Splatting\cite{yu2024mip}, 3DGS-MCMC~\cite{kheradmand20243d} and \mbox{Scaffold-GS~\cite{lu2024scaffold}}. Since our method emphasizes the initialization stage, we include RAIN-GS~\cite{jung2024relaxing}. Notably, the mean values for Scaffold-GS and 3DGS-MCMC changed significantly, as they originally reported results for only 7 of the 9 Mip-NeRF360 scenes. Additionally, we report results for Scaffold-GS trained with the same resolution settings as 3DGS, which were not included in the original paper. We compare these models against our model with full $30000$-step convergence, pruning enabled, densification disabled denoted as \mbox{\textit{EDGS + 3DGS}} in~\cref{tab:merged_comparison}. 

To evaluate speed and efficiency, we compare our model without densification, stopped at $5000$  and $10000$ steps (\textit{EDGS + 3DGS 5K} and \textit{10K}) in~\cref{tab:efficciency_comparison}) against the fastest competitive methods: EAGLES~\cite{girish2023eagles}, 3DGS-LM~\cite{hollein20243dgs}, Taming 3DGS~\cite{mallick2024taming}, gsplat~\cite{ye2024gsplatopensourcelibraryGaussian}, and MiniSplatting~\cite{Fang2024MiniSplattingRS}.
\subsection{Quantitative Evaluations}
\label{sec:quant}
We evaluate EDGS across three aspects: reconstruction quality, training efficiency, and compatibility with existing methods. 
As shown in~\cref{tab:merged_comparison}, our model achieves the best or second-best results across all three metrics, while maintaining comparable training time and Gaussian count, all without any densification. 
When trained for only 5K steps, EDGS matches the performance of efficiency-focused methods while converging faster~(\cref{tab:efficciency_comparison}). 
Moreover, EDGS can be seamlessly integrated with existing Adaptive Density Control (ADC) methods by using it as an initialization. As shown in~\cref{tab:ours_plus_adc}, this integration consistently improves reconstruction quality across all evaluated ADC variants on the Mip-NeRF360~\cite{barron_mip-nerf_2022} dataset. Since EDGS itself does not perform densification, we intentionally initialize with fewer Gaussians in this experiment to allow ADC methods to further refine the scene. All ADC variants use identical initialization parameters for fairness. Despite the additional densification, total training time remains comparable or lower, as fewer Gaussians are introduced overall and less computation is spent on densification. Finally, unlike prior 3DGS approaches that rely heavily on densification and degrade significantly without it, EDGS remains stable and achieves strong reconstruction quality even without it, see~\cref{tab:densification}.

\subsection{Qualitative Evaluations}
\label{sec:qual}
\begin{figure*}
  \centering
   \includegraphics[width=0.99\textwidth]{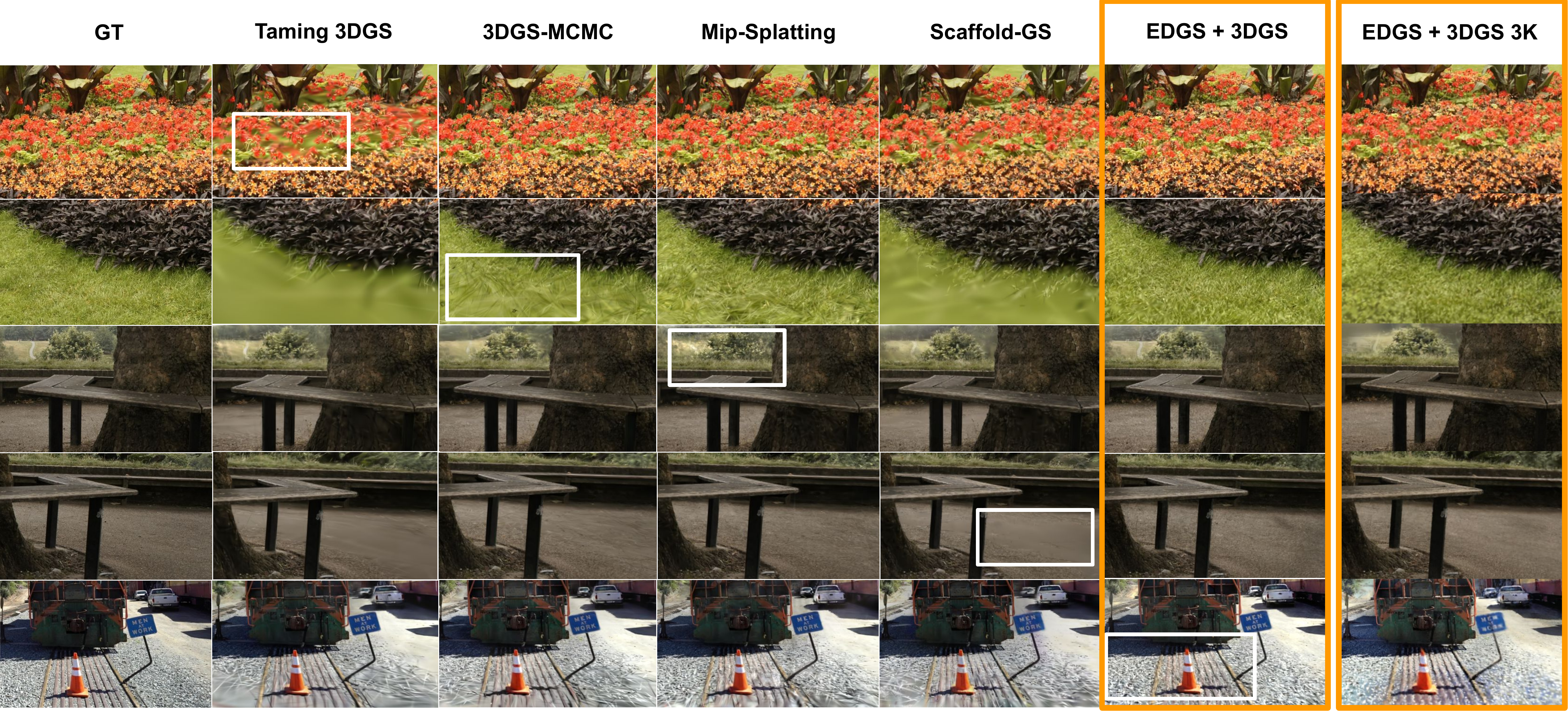}
   \caption{
   Qualitative comparison on \textit{flowers} and \textit{treehill} scenes from Mip-NeRF360~\cite{barron_mip-nerf_2022} and \textit{train} scene from Tanks\&Temples~\cite{knapitsch2017tanks}. EDGS produces sharper and more consistent reconstructions across diverse scenes. Even when trained for only 3000 steps, it matches state-of-the-art perceptual quality. For this visualization, we crop regions of interest. The full renderings are provided in~\cref{sec:sup:visual}.}
  \label{fig:qualitative_comparison}
\end{figure*}

In~\cref{fig:qualitative_comparison}, our approach shows clear improvements over other methods across all datasets. The examples show that EDGS excels not only in high-frequency regions, such as small stones near railroad tracks, grass, or concrete textures, but also in capturing fine details like flower stems (first row) and distant elements like roads (third row). Other models often fail to accurately reconstruct these details, either blurring them or introducing high-frequency artifacts. EDGS dense initialization ensures a Gaussian splat is placed at every meaningful location, enabling precise and detailed reconstruction. We also provide crops for our model, stopped at 3000 steps, showing that we achieve comparable perceptual quality much faster than other methods.

\begin{table}[t]
    \centering
    \adjustbox{max width=0.85\columnwidth }{
    \begin{tabular}{l@{\hskip 10mm}c@{\hskip 10mm}ccc}
        \toprule
        \multirow{2}{*}{\textbf{Method}}  & \multirow{2}{*}{%
  \begin{tabular}[c]{@{}c@{}}%
    {\small \textbf{Densifica-}}\\
    {\small \textbf{tion free}} \\
  \end{tabular}%
}  & \multirow{2}{*}{\textbf{PSNR$\uparrow$}} & \multirow{2}{*}{\textbf{SSIM$\uparrow$}} & \multirow{2}{*}{\textbf{LPIPS$\downarrow$}} \\
& & & & \\
        \midrule
                   & {\color{ourgreen}\cmark} & 25.60 & 0.709 & 0.367 \\
                                         \multirow{-2}{*}{3DGS}    & {\color{ourred}\xmark} & 27.49 & 0.816 & 0.215 \\
        \midrule
        \multirow{2}{*}{\textbf{EDGS}}   & {\color{ourgreen}\cmark}   & \underline{28.02} & \textbf{0.839} & \underline{0.141} \\
            & {\color{ourred}\xmark} & \textbf{28.08} & \underline{0.831} & \textbf{0.140} \\
        \bottomrule
    \end{tabular}
    }
    \caption{In contrast to 3DGS, EDGS does not require densification and only marginally improves when densification is applied.}
    \label{tab:densification}
\end{table}

\subsection{Ablation Studies}
\label{sec:abl}

\paragraph{Matching Algorithm Comparison}  We evaluate image matching methods $\mathcal{M}$ for initializing splats. See~\cref{tab:matching_algorithms,} for a comparison on the Mip-NeRF360 dataset. Throughout this paper, we use RoMa~\cite{edstedt_roma_2023} as our matching algorithm,  but we also experiment with LoFTR~\cite{sun2021loftr}, DKM~\cite{edstedt2023dkm}, and RAFT~\cite{Teed2020RAFTRA}. All methods except RAFT achieve comparable performance; RAFT struggles due to its primary design for optical flow between consecutive video frames, where viewpoint differences are minimal.

\begin{table}[t]
    \centering
    \adjustbox{max width=0.95\columnwidth }{
    \begin{tabular}{l@{\hskip 10mm}c@{\hskip 10mm}ccc}
        \toprule
        \multirow{1}{*}{\textbf{Method}} & \multirow{1}{*}{\textbf{Matching}} & \multirow{1}{*}{\textbf{PSNR$\uparrow$}} & \multirow{1}{*}{\textbf{SSIM$\uparrow$}} & \multirow{1}{*}{\textbf{LPIPS$\downarrow$}}     \\
        \midrule
         3DGS          &  & 27.49 & 0.816 & 0.215 \\
        \multirow{4}{*}{\textbf{EDGS}}          & LoFTR~\cite{sun2021loftr}      & 27.79 & 0.818 & 0.179 \\
                  & DKM~\cite{edstedt2023dkm}    & 27.81 & 0.831 & 0.192 \\
               & RAFT~\cite{Teed2020RAFTRA}     & 26.90 & 0.803 & 0.201 \\
           & RoMa~\cite{edstedt_roma_2023}    & \textbf{28.02} & \textbf{0.839} & \textbf{0.141} \\
        \bottomrule
    \end{tabular}
    }
    \caption{EDGS can leverage various dense feature matching algorithms and consistently achieves high reconstruction scores across them, even without using densification.}

   \label{tab:matching_algorithms}
\end{table}

\begin{figure}
    \captionsetup{type=figure}
       \includegraphics[width=.49\linewidth]{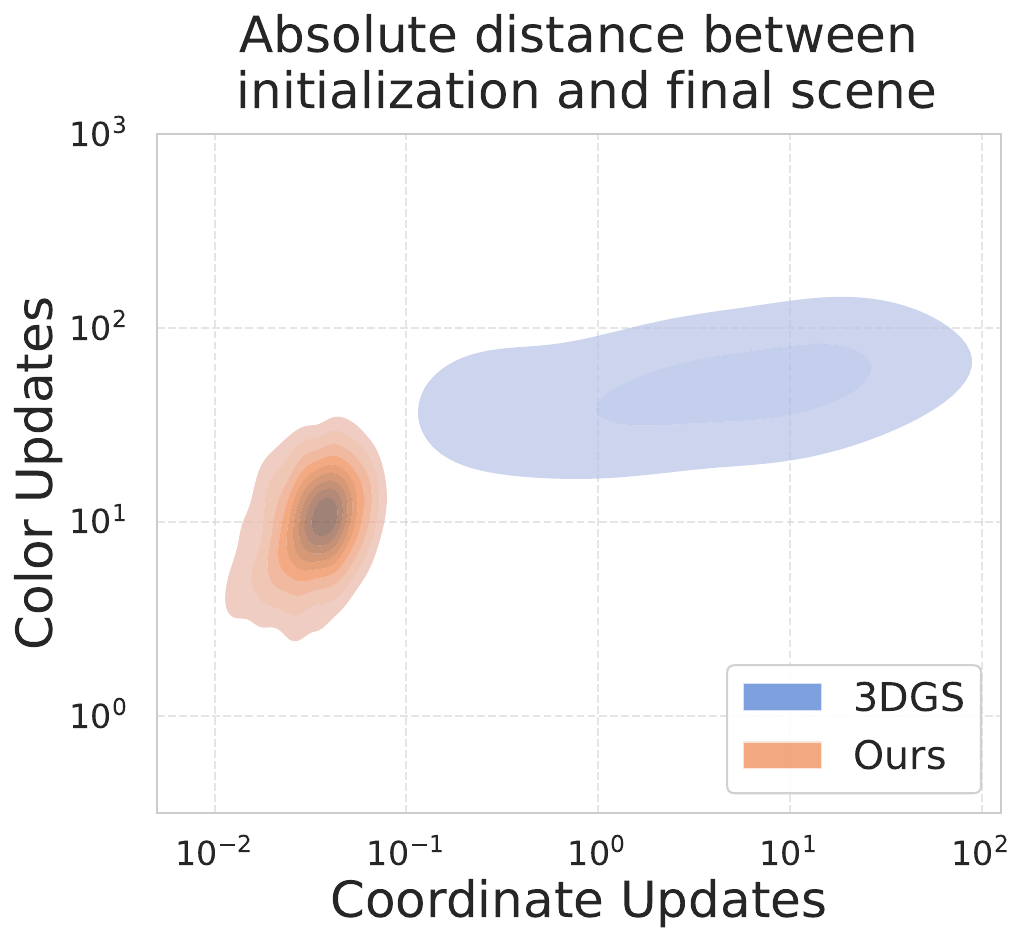} 
        \includegraphics[width=.49\linewidth]{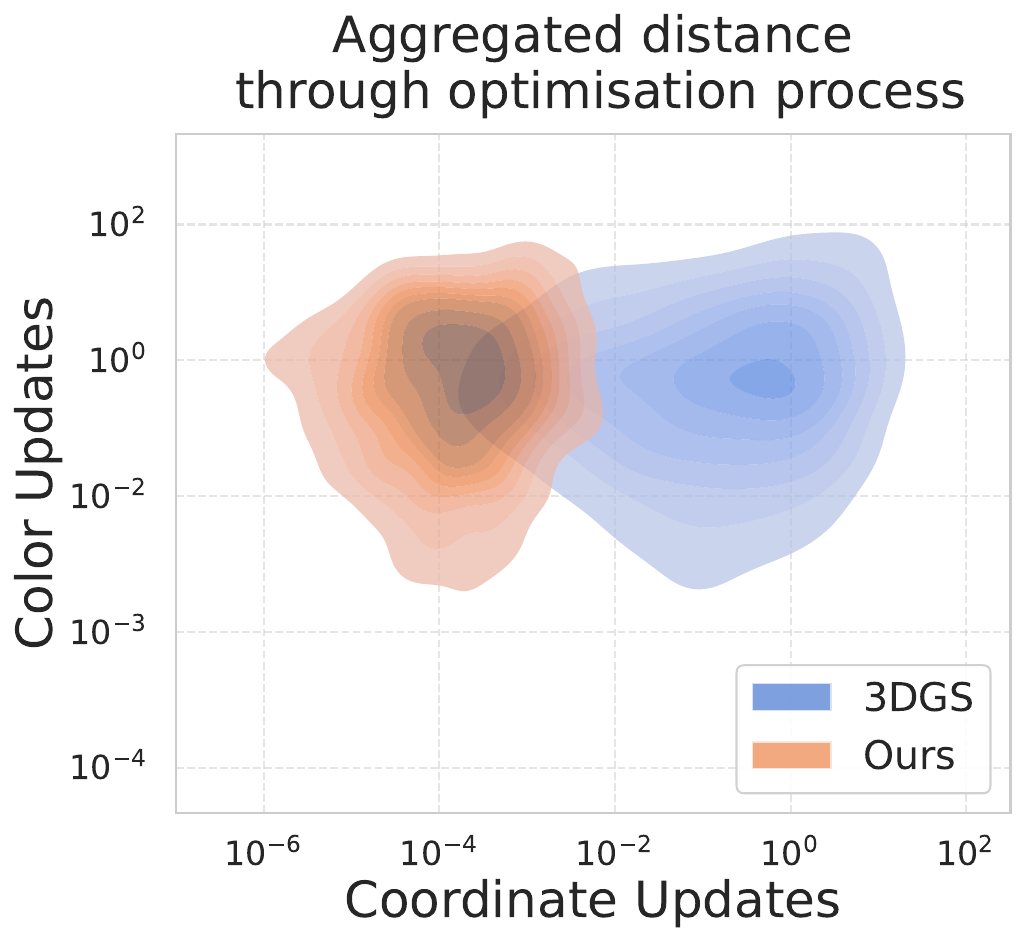} 
  \caption{
 Distributions of Gaussian parameters change in color/coordinate space throughout training. Our EDGS not only initializes closer to the solution (left) but also requires significantly fewer adjustments (right) during optimisation process, leading to faster and more stable convergence. }
  \label{fig:colorcoor_adjustment}
\end{figure}

\paragraph{Gaussian Motion and Convergence} 
We study the parameters dynamics of each Gaussian during optimization. \cref{fig:colorcoor_adjustment} presents the start-to-finish displacement and full motion trajectory length. Namely, we analyze how Gaussian coordinates and color parameters evolve during the optimization process by measuring two key distributions.
Let \( \bm{g}_i(t) \) denote the state of Gaussian \( \bm{g}_i \) at optimization step \( t \) for  $ i \in \{1, \dots, N\}$. The first distribution captures the \textit{displacement}, defined as:
\begin{equation}
    \begin{pmatrix}
        \| \bm{g}_i^c(0) - \bm{g}_i^c(T) \|_2 \\
        \| \bm{g}_i^x(0) - \bm{g}_i^x(T) \|_2
    \end{pmatrix} \in \mathbb{R}^2,
\end{equation}
and second measures the full \textit{trajectory length}:

\begin{equation}
    \begin{pmatrix}
        \sum\nolimits^{T}_{t=0} \| \bm{g}_i^c(t) - \bm{g}_i^c(t+1) \|_2 \\
        \sum\nolimits^{T}_{t=0} \| \bm{g}_i^x(t) - \bm{g}_i^x(t+1) \|_2
    \end{pmatrix} \in \mathbb{R}^2,
\end{equation}
where $T$ denotes the number of optimization steps. EDGS significantly reduces the final coordinate displacement, as Gaussians are initialized closer to surfaces, requiring fewer adjustments. Compared to 3DGS, our model reduces the coordinate displacement by 50 times and the coordinate trajectory length by 30 times. The color trajectory length also decreases, though less dramatically, by approximately a factor of two, as small oscillations remain along the trajectory. Visualizations of Gaussian motion are provided in~\cref{sec:sup:Gaus_motion}.

\begin{table}[t]
    \centering
    \adjustbox{max width=0.99\columnwidth}{
    \begin{tabular}{lcccccc}
        \toprule
        \textbf{Method} & $\mathbf{p}^{ij}_{\text{corr}}$ & $\mathbf{p}^{ij}_{\text{proj}}$ & \textbf{SH Init.} & \textbf{PSNR$\uparrow$} & \textbf{SSIM$\uparrow$} & \textbf{LPIPS$\downarrow$} \\
        \midrule
        \textbf{EDGS (full)}     & {\color{ourgreen}\cmark} & {\color{ourgreen}\cmark}            & {\color{ourgreen}\cmark} & \textbf{28.02} & \underline{0.839} & \textbf{0.141} \\
        w/o SH init.             & {\color{ourgreen}\cmark} & {\color{ourgreen}\cmark}            & {\color{ourred}\xmark}   & 27.80 & \textbf{0.840} & 0.175 \\
        w/o $\mathbf{p}^{ij}_{\text{proj}}$ & {\color{ourgreen}\cmark} & {\color{ourred}\xmark}   & {\color{ourred}\xmark}   & 27.72 & 0.830 & 0.179 \\
        w/o $\mathbf{p}^{ij}_{\text{corr}}$ & {\color{ourred}\xmark}   & {\color{ourgreen}\cmark} & {\color{ourred}\xmark}   & 27.55 & 0.829 & 0.197 \\
        baseline     & {\color{ourred}\xmark}   & {\color{ourred}\xmark}   & {\color{ourred}\xmark}   & 27.43 & 0.822 & 0.202 \\
        \bottomrule
    \end{tabular}
    }
    \caption{Ablation study of EDGS components. Combining both sampling distributions with spherical harmonics initialization(SH Init) yields the best reconstruction quality.}
    \label{tab:hyperparam_abl}
\end{table}

\paragraph{Sampling distribution} 
As defined in \cref{eq:methods:sampling}, the sampling distribution $\mathbf{p}^i$ combines two terms: the correspondence-based distribution $\mathbf{p}^{ij}{\text{corr}}$, which reflects matcher confidence, and the geometry-based distribution $\mathbf{p}^{ij}_{\text{proj}}$, which penalizes re-projection errors. These components complement each other: confidence alone captures semantic reliability but cannot enforce geometric consistency, whereas the re-projection term removes mismatched or unstable correspondences. To maintain spatial coverage, we perform uniform sampling over confidence-thresholded matches $\mathbf{c}^{ij}$, preventing bias toward high-confidence but spatially clustered regions. The combination of both terms yields a balanced, geometry-aware initialization that consistently outperforms using either component in isolation (\cref{tab:hyperparam_abl}).
\begin{figure}
  \centering
   \includegraphics[width=1\columnwidth]{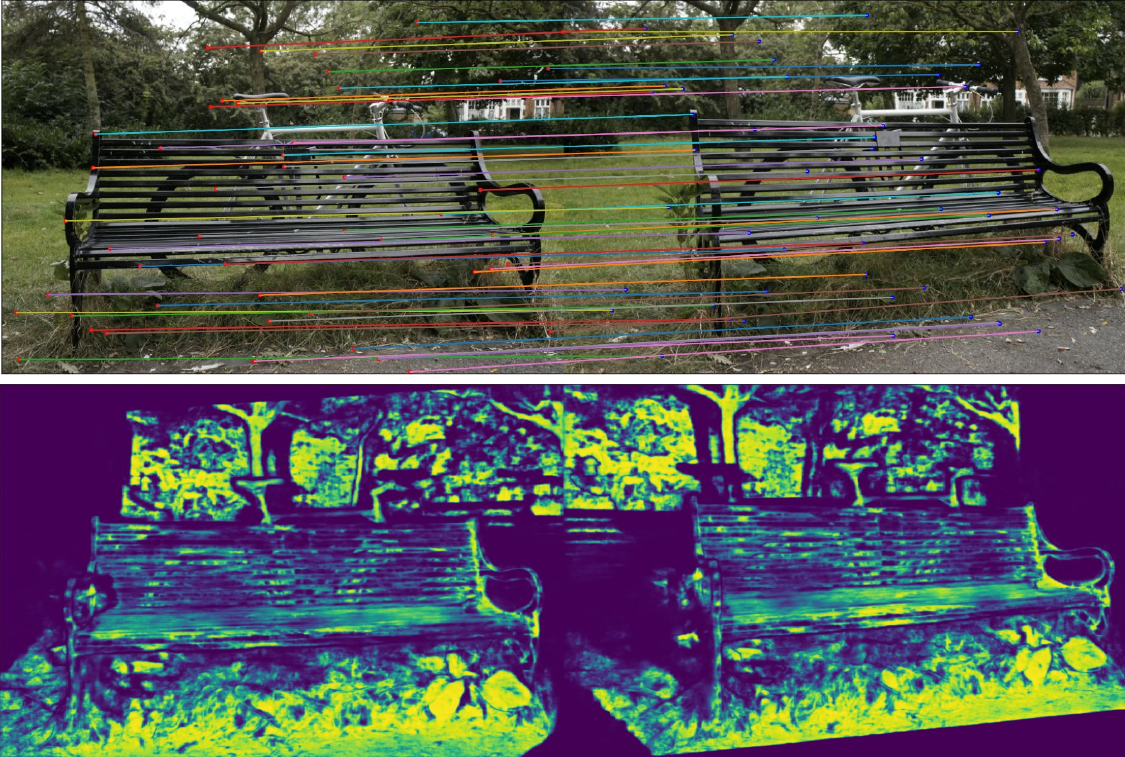}
   \caption{
   Dense correspondences for a pair of images from \textit{bicycle} scene. The top row shows matched keypoints, and the bottom row visualizes the confidence of correspondences in the neighboring image. The matching model $\mathcal{M}$ is RoMa~\cite{edstedt_roma_2023}. Correspondence confidence is not uniform across the scene.
   }
  \label{fig:keypoints_confidence}
\end{figure}

To assess the importance of the re-projection term, we replace the right-hand side of \cref{eq:methods:sampling} with $\max_{j \in \mathbb{I}i} \mathbf{p}^{ij}_{\text{corr}}(k)$. Conversely, removing $\mathbf{p}^{ij}_{\text{corr}}$ entirely is infeasible as we still require match locations, but we can sample points proportionally to confidence by setting $\mathbf{p}^{ij}_{\text{corr}} \propto \mathbf{c}^{ij}$. As shown in~\cref{tab:hyperparam_abl}, both modifications degrade performance, with the confidence-only variant having the strongest negative effect. This is expected: sampling in proportion to $\mathbf{c}^{ij}$ breaks uniform spatial coverage and concentrates splats around high-confidence edges or boundaries, leading to poorer initialization. Disabling both geometric and confidence cues further reduces accuracy and produces visibly more floaters.  In~\cref{fig:keypoints_confidence}, we visualize a set of keypoints extracted from a single pair of images. The results highlight that we need to sample keypoints from the image more uniformly, rather than focusing solely on keypoints with high confidence.

\begin{figure}
  \centering
   \includegraphics[width=.95\columnwidth]{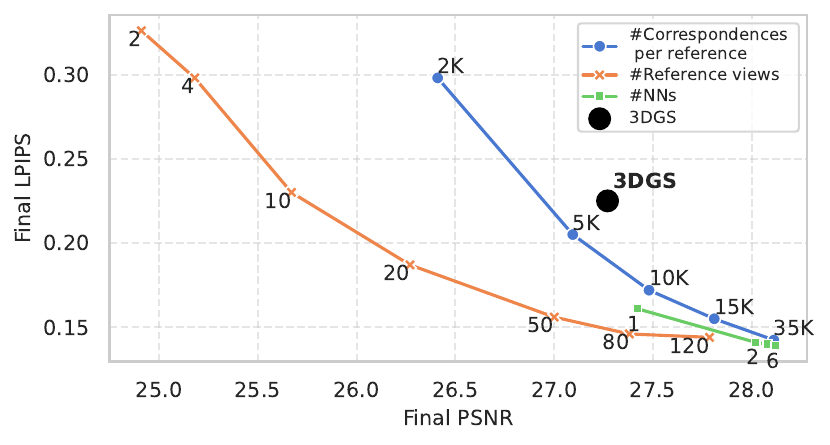}
   \caption{
   PSNR and LPIPS curves show saturation as we independently increase any of three parameters: number of reference views, number of nearest-neighbor, and number of sampled correspondences per reference view.  3DGS is provided for reference. 
   }  \label{fig:ablation_num_reference_num_matches}
\end{figure}

\paragraph{Hyperparameter Sensitivity}
We analyze how key hyperparameters, namely the number of reference views, the number of sampled correspondences per view, and the number of nearest neighbor views used for matching, affect reconstruction quality. As shown in~\cref{fig:ablation_num_reference_num_matches}, increasing any of these parameters improves results until a saturation point, after which gains become marginal. Following this behavior, we use up to $180$ reference views (limited by scene size), match each with $2$ nearest neighbors to obtain $\mathbf{p}^i$, from which we sample $20\text{k}$ correspondences. 
See~\cref{sec:sup:impact_nn} for details. 

\paragraph{Spherical harmonics} We further evaluate the effect of EDGS spherical harmonics initialization in~\cref{tab:hyperparam_abl}. This initialization accelerates convergence for challenging Gaussians and improves the modeling of view-dependent effects, resulting in significantly lower LPIPS scores. We also observe that the benefits are more pronounced for indoor scenes, where complex lighting and reflections pose greater challenges, compared to outdoor environments.

\begin{figure}
  \centering
   \includegraphics[width=.99\columnwidth]{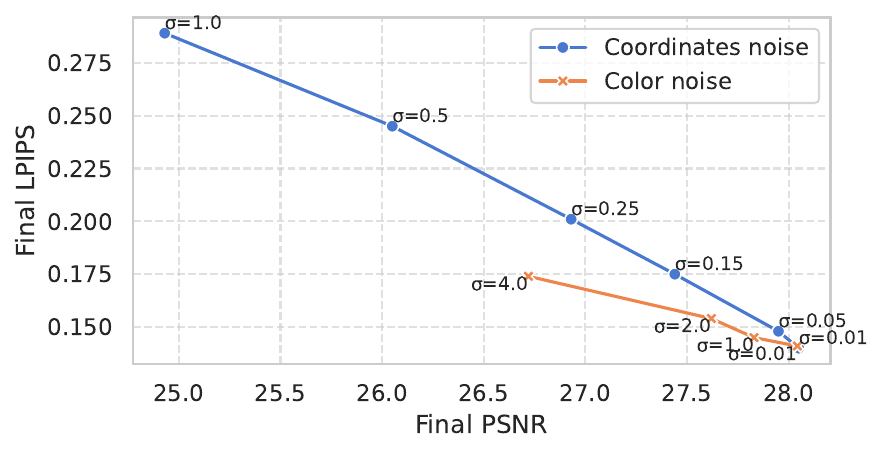}
   \caption{
   Robustness under initialization noise. Note that the noise scale $\sigma$ is larger for color. EDGS is robust to inaccuracies in the initialization, maintaining high reconstruction quality.
   }
  \label{fig:ablation_noise}
\end{figure}

\paragraph{Robustness to noise}
\label{sec:robustness_to_noise}
EDGS exhibits strong resilience to imperfections in the initial correspondences, which may arise from triangulation inaccuracies or suboptimal matches produced by $\mathcal{M}$. To quantitatively assess this robustness, we perturb the initialized Gaussian parameters with additive Gaussian noise $\epsilon \sim \mathcal{N}(0,\sigma)$ applied independently to either spatial coordinates or color values. We then evaluate reconstruction quality across varying noise levels $\sigma$. Formally, noise $\epsilon$ is injected into the color parameters $\bm{g}_i^c$ and coordinate parameters $\bm{g}_i^x$ of the initialized Gaussians. This design isolates the influence of each parameter type on convergence.

As shown in~\cref{fig:ablation_noise}, increasing coordinate or color noise leads to gradual degradation in PSNR and LPIPS. Remarkably, EDGS maintains substantially higher robustness to color perturbations than to spatial ones, supporting our claim that the method’s primary advantage lies in reducing unnecessary Gaussian movement during optimization. Even under moderate coordinate noise, overall reconstruction quality remains stable, demonstrating the inherent regularization of our initialization. Notably, our method remains stable even with small amounts of added noise, likely because the initialization itself is already inherently noisy, as shown in~\cref{fig:inits}. All experiments are conducted on the Mip-NeRF360 dataset.
In~\cref{fig:ablation_noise_sup}, we visualized initialization for scene \textit{garden}, which was noised with different scales for both coordinates (first row) and colors (second row).

\paragraph{Different Initializations} 
\begin{table}
    \centering
    \adjustbox{max width=0.99\columnwidth }{
    \begin{tabular}{llcccc}
        \toprule
        \multirow{2}{*}{\textbf{Method}} & \multicolumn{1}{c}{\textbf{Init}} & \multicolumn{1}{c}{\textbf{Densif.}} & \multirow{2}{*}{\textbf{PSNR$\uparrow$}} & \multirow{2}{*}{\textbf{SSIM$\uparrow$}} & \multirow{2}{*}{\textbf{LPIPS$\downarrow$}} \\
        & \multicolumn{1}{c}{\textbf{type}} & \multicolumn{1}{c}{\textbf{free}} & & & \\
        \midrule
        \multirow{4}{*}{3DGS}                  & Random  & {\color{ourred}\xmark} & 22.19 & 0.704 & 0.313 \\
                   & COLMAP  & {\color{ourred}\xmark} & 27.49 & 0.816 & 0.215 \\
           & Depth~\cite{gui_depthfm_2024}   & {\color{ourgreen}\cmark}   & 26.99 & 0.810 & 0.202 \\
            & Depth~\cite{gui_depthfm_2024}   & {\color{ourred}\xmark} & 27.18 & 0.819 & 0.197 \\
        \rowcolor{gray!10} VGGT-X~\cite{liu2025vggt}   & VGGT~\cite{wang2025vggt} & {\color{ourred}\xmark} &  26.40 & 0.782 & 0.177 \\
        \textbf{EDGS}  & EDGS & {\color{ourgreen}\cmark}   & \textbf{28.02} & \textbf{0.839} & \textbf{0.141} \\
        \bottomrule
    \end{tabular}
    }
    \caption{Qualitative comparison for different initialization strategies and densification. Checkmarks in the \textbf{Densif. free} column indicate models not using 3DGS densification~\cite{kerbl20233d}. EDGS consistently achieves superior results across all metrics.}
    \label{tab:init}
\end{table}

We study now different ways to initialize Gaussians for 3DGS. The first one is random noise, but this setup fails to achieve the performance of SFM-based initialization, further highlighting the critical role of proper initialization. Beyond matching-based initialization, we evaluate a depth-based strategy using DepthFM~\cite{gui_depthfm_2024}. For each reference view, we uniformly sample 20,000 pixels and backproject them to 3D using the predicted depth. However, monocular depth estimates often suffer from scale inconsistencies across views, leading to lower reconstruction quality. While DepthFM combined with densification outperforms the baseline 3DGS, it still falls short of our matching-based approach. We also compare with a neural initialization method that jointly estimates Gaussian positions and camera parameters~\cite{liu2025vggt}; however, it does not reach the same reconstruction quality as EDGS. See~\cref{tab:init} for quantitative comparison on the Mip-NeRF360. We additionally include, in~\cref{sec:supp_init_baselines}, further simpler initialization baselines.

\begin{figure}[ht!]
  \centering
   \includegraphics[width=1.0\columnwidth]{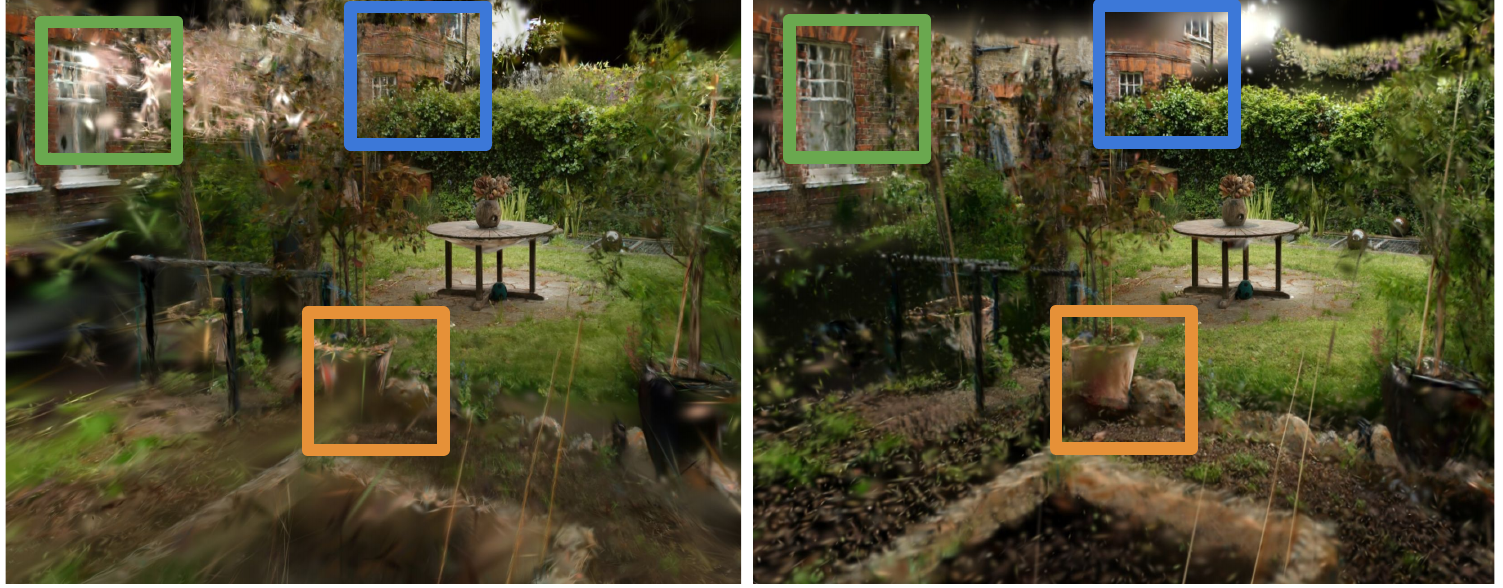}
   \caption{Extreme viewpoint rendering.
   EDGS~(right) better preserves details and reduces stretched Gaussians when rendering from viewpoints far outside the training set compared to the 3DGS~(left). This results in a more consistent distribution and improved quality, especially in challenging regions like the building and flower pot.
   }
   \vspace{6pt}
  \label{fig:extreme_viewpoint}
\end{figure}

\paragraph{Extreme Viewpoint Rendering}
EDGS effectively handles extreme viewpoint variations, outperforming the baseline when rendering from camera angles far outside the training set. As shown in~\cref{fig:extreme_viewpoint}, our dense initialization prevents the need for stretching small Gaussians to compensate for pixel loss at a distance, resulting in a more stable and accurate reconstruction. As visualized for \textit{garden} scene from the Mip-NeRF360 dataset, our method avoids large Gaussians and exhibits less noise compared to the competing approach.

\paragraph{Additional experiments and details}
Implementation details are provided in~\cref{sec:sup:implement_det}, with extra visualizations in~\cref{sec:sup:visual}. For the table of notation see~\cref{sec:sup:notation}. We also evaluate our method in a sparse view setting  in~\cref{sec:sup:few_view} and show applicability of our approach to this setting as well.
\section{Conclusion}
We introduce a new initialization strategy for 3D Gaussian Splatting that removes the need for iterative densification. The approach relies on carefully sampling 2D correspondences that are both geometrically consistent and provide uniform, dense coverage of the scene. 

Our method reaches state-of-the-art performance without any densification and matches efficiency-oriented methods with substantially fewer optimization steps. Moreover, EDGS functions as a plug-and-play initialization for adaptive density control techniques, improving reconstruction quality without increasing training time or Gaussian count, making it a practical and broadly applicable enhancement for 3D reconstruction pipelines.

\section*{Acknowledgement}
We thank Stefan Baumann, Felix Krause, and Pingchuan Ma for feedback, proofreading, and helpful discussions.
This project has been supported by the German Federal Ministry for Economic Affairs and Climate Action within the project ``NXT GEN AI METHODS – Generative Methoden für Perzeption, Prädiktion und Planung'', the bidt project KLIMA-MEMES, Bayer AG, the project “GeniusRobot” (01IS24083), funded by the Federal Ministry of Education and Research (BMBF). The authors gratefully acknowledge the Gauss Center for Supercomputing for providing compute through the NIC on JUWELS at JSC and the HPC resources supplied by the Erlangen National High Performance Computing Center (NHR@FAU funded by DFG project 440719683) under the NHR project JA-22883. Further, we thank Owen Vincent for continuous technical support.
\FloatBarrier
{
    \small
    \bibliographystyle{ieeenat_fullname}
    \bibliography{main}
}


\clearpage
\appendix
\renewcommand\thefigure{A\arabic{figure}}
\renewcommand\thetable{A\arabic{table}}
\renewcommand\theequation{A\arabic{equation}}
\setcounter{equation}{0}
\setcounter{table}{0}
\setcounter{figure}{0}
\setcounter{page}{1}

\newpage

\maketitlesupplementary

\section{Implementation details}
\label{sec:sup:implement_det}

\paragraph{Evaluation protocol}
Following standard practice in 3DGS-based reconstruction, every 8th camera view is used for testing. For the Mip-NeRF360 dataset, we follow the original 3DGS protocol~\cite{kerbl20233d} and downsample outdoor scenes by a factor of four and indoor scenes by a factor of two. Other datasets are used at their original resolution.

\paragraph{Initialization}
We initialize the scene using up to $180$ reference views. For each reference view $I_i$, we select its two nearest neighbors based on camera-pose proximity and compute dense correspondences using RoMa~\cite{edstedt_roma_2023}. Each forward matching pass takes $0.21\,\text{s}$ on an NVIDIA A100 GPU. For every correspondence, we compute the triangulated 3D point and evaluate its reprojection error in both participating views. We keep only matches with confidence above $\tau_{\text{corr}} = 0.05$ and reprojection error below $\tau_{\text{proj}} = 0.01$ (in NDC units). From the resulting geometrically consistent set, we sample $20\text{K}$ correspondences per reference view according to our distribution $\mathbf{p}_i$. For each sampled point, we estimate the spherical harmonics coefficients, initialize the color from the corresponding pixel in $I_i$, and set its initial scale proportional to its distance from the reference camera. 

For the default setting (180 reference views, 2NN, 20K correspondences/view), full-scene initialization takes $\sim120\,\text{s}$ end-to-end on an A100. Dense matching dominates this cost: $180{\times}2$ forward passes at $0.21\,\text{s}$ each account for $\approx76\,\text{s}$. Triangulation and SH estimation are run iteratively per view and take $\sim11\,\text{s}$ and $\sim15\,\text{s}$, respectively; the remaining time is spent on reprojection-based filtering, data preparation, and splat instantiation. We use a single CPU core for auxiliary preprocessing, and peak GPU memory during initialization is $\sim15$\,GB.

\paragraph{Spherical harmonics initialization}
As outlined in the main manuscript (Sec.~\ref{sec:sh}), we estimate spherical harmonics (SH) coefficients for each sampled Gaussian using its available multi-view color observations. We provide additional implementation details here.

Each Gaussian typically has only two usable observations (one from the reference view and one from its nearest neighbor), making the SH estimation problem underdetermined.
Despite the limited observations, we formulate the full SH coefficient matrix $\hat{\mathbf{H}}_k \in \mathbb{R}^{16 \times 3}$ (degree $l=3$) and initialize all coefficients except the first component (index~0) using the pseudoinverse solution $\mathbf{Y}_k^{+}\mathbf{O}_k$.
The first component is initialized directly from the reference-view color at pixel $(u_k^i, v_k^i)$ in $I^i$, ensuring that the initial appearance is consistent with the reference image while still providing a stable estimate for higher-order coefficients.

Following the standard 3DGS optimization protocol, we progressively unfreeze the SH coefficients during the optimization process to ensure a fair comparison with prior work. The color component and the first four spherical harmonics coefficients are optimized from initialization, with each subsequent coefficient progressively unfrozen every 1,000 iterations.

\paragraph{Optimization}
After initialization, we employ the standard 3DGS optimization schedule, disabling densification and omitting gradient aggregation for detecting under-reconstructed regions in order to isolate the effect of initialization. All models are trained for the same number of iterations as competing methods (30000 steps) unless explicitly stopped early at $5\text{K}$ or $10\text{K}$ steps (\textit{EDGS + 3DGS 5K} and \textit{EDGS + 3DGS 10K}). All experiments were conducted on an NVIDIA A100 (80\,GB). Our method requires at most $15\,\text{GB}$ of GPU memory.

\paragraph{Integration with ADC methods}
When combining EDGS with adaptive density-control methods, we enable densification but maintain the final Gaussian count by initializing with fewer points. Specifically, we use $140$ reference views, $8.5\text{k}$ sampled correspondences per view, and two nearest neighbors, which yields initialization sizes comparable to those produced by the other ADC strategies.

\section{Sparse-view setting}
\label{sec:sup:few_view}
\begin{table}
\centering
\adjustbox{max width=0.99\columnwidth}{
\begin{tabular}{lc@{\hskip 2mm}c@{\hskip 2mm}c@{\hskip 7mm}c@{\hskip 2mm}c@{\hskip 2mm}c}
\toprule
 \multirow{2}{*}{\textbf{Methods}} & \multicolumn{3}{c}{\textbf{12-view}} & \multicolumn{3}{c}{\textbf{24-view}} \\
\cmidrule(lr){2-7} 
 &\textbf{SSIM $\uparrow$} & \textbf{PSNR $\uparrow$} & \textbf{LPIPS $\downarrow$} & \textbf{SSIM $\uparrow$}  & \textbf{PSNR $\uparrow$} & \textbf{LPIPS $\downarrow$}  \\ \midrule

Mip-NeRF 360~\cite{barron_mip-nerf_2022} & 0.432 & 17.73 & 0.520 & 0.530 & 19.78 & 0.431  \\
RegNeRF~\cite{niemeyer2022regnerf} & 0.437 & 18.84 & 0.544 & 0.546 & 20.55 & 0.398  \\
SparseNeRF~\cite{wang2023sparsenerf} & 0.395 & 17.44 & 0.609 & 0.600 & 21.13 & 0.389  \\
3DGS~\cite{kerbl20233d} & 0.499 & 17.49 & 0.431 & 0.588 & 19.93 & 0.401  \\
SparseGS~\cite{xiong2023sparsegs} & \underline{0.577} & \textbf{19.37} & \underline{0.398} & \textbf{0.713}  & \textbf{23.02} & \underline{0.290}  \\
\midrule
\textbf{EDGS + 3DGS} & \textbf{0.594} & \underline{18.96} & \textbf{0.388} & \underline{0.699} & \underline{22.25} & \textbf{0.289}  \\
\bottomrule
\end{tabular}
}
\caption{Quantitative results on the Mip-NeRF360~\cite{barron_mip-nerf_2022} dataset under 12- and 24-view training settings. Although EDGS is not designed for sparse-view reconstruction, it performs on par with specialized baselines and in some cases surpasses SparseGS~\cite{xiong2023sparsegs} that leverage diffusion-based score distillation losses.}
\label{tab:few_view}
\end{table}

We evaluate our method in sparse-view settings following the protocol established by SparseGS~\cite{xiong2023sparsegs}. Experiments are conducted on seven scenes from the Mip-NeRF360 dataset, excluding \textit{flowers} and \textit{treehill}, to ensure a fair comparison with prior work. For each scene, we reserve every eighth image as a test view and uniformly sample either 12 or 24 of the remaining images as the training set.

Training schedules match previous standards: 10k iterations for the 12-view setup and 30k iterations for the 24-view setup. In this experiment, we disable densification for EDGS to focus on the effect of adding our initialization. For each reference view, we sample 50{,}000 correspondences and treat all training views as reference views, selecting the two nearest neighboring views for correspondence matching.

We compare against established baselines for sparse reconstruction, including RegNeRF~\cite{niemeyer2022regnerf}, SparseNeRF~\cite{wang2023sparsenerf}, the original 3DGS method~\cite{kerbl20233d}, and the state-of-the-art sparse 3DGS variant, SparseGS~\cite{xiong2023sparsegs}.

Even without densification, EDGS delivers performance competitive with methods specifically tailored for sparse-view reconstruction. As demonstrated in~\cref{tab:few_view}, the proposed initialization provides reliable geometric alignment and stable optimization, enabling EDGS to match or exceed methods that rely on additional regularization or learned priors~\cite{xiong2023sparsegs} in handling limited supervision. These results indicate that a strong correspondence-based initialization alone can substantially improve the quality of 3DGS reconstruction in sparse-view scenarios. This also suggests that densification is not essential for achieving robust performance when the initialization effectively captures scene geometry given only limited number of views.

\section{Nearest Neighbors}
\label{sec:sup:impact_nn}
\begin{figure}[ht!]
  \centering
   \includegraphics[width=.99\columnwidth]{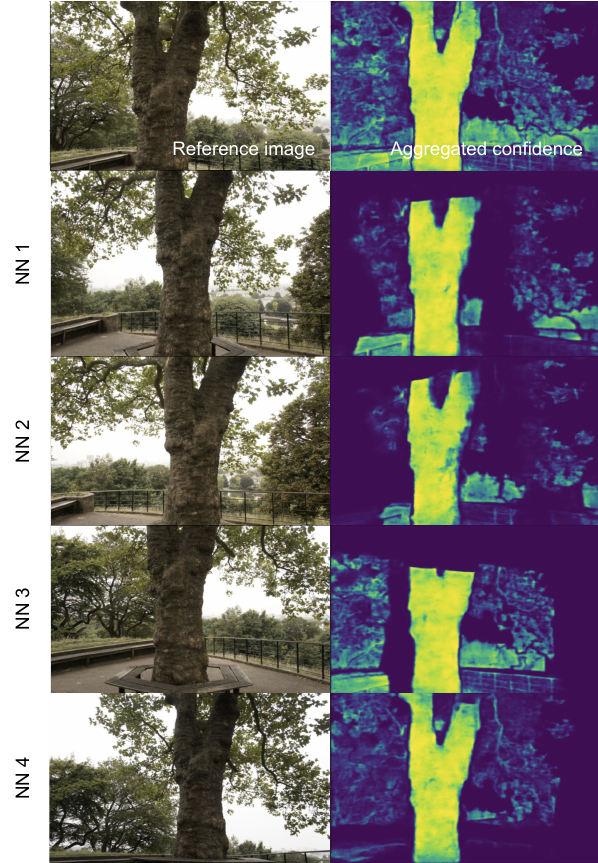}
   \caption{
   Visualization of correspondence extraction from multiple nearest neighbors for reference image $I^i$. The top-left picture shows the reference view. Each subsequent row displays a neighboring ground-truth image (left), ordered by camera proximity, and its corresponding matching confidence map $\mathbf{c}^{ij}$ (right). The top-right picture shows the aggregated confidence map $\mathbf{c}^i =
     \max_{j \in \mathbb{I}_i} \mathbf{c}^{ij}$, formed by combining scores from all neighbors. Aggregation provides denser and more uniform coverage of the reference frame. Example shown for the \textit{treehill} scene from Mip-NeRF360.}
  \label{fig:sup_corr_densities}
\end{figure}

Ensuring full scene coverage requires sampling from all regions of each reference view. However, a single neighboring view typically overlaps with only a subset of the reference image, meaning that reliable correspondences exist only for those shared regions. To cover the entire reference view with reliable matches, we therefore aggregate correspondences from multiple nearest neighbors. Using a larger number of reference views further ensures that all parts of the scene receive initial splats.
In~\cref{fig:sup_corr_densities}, we visualize the contribution of individual neighbors. The reference image $I^i$ is shown in the top-left. Rows 2–5 show, for each nearest camera $j$(sorted by distance to the reference view camera), the corresponding ground-truth view (left) and confidence map $\mathbf{c}^{ij}$ (right), indicating which pixels in $I^i$ were matched to that neighbor. Each neighbor covers only a subset of the reference view, which motivates aggregating confidence values on a per-pixel basis. We therefore define the aggregated confidence map for a reference view $I_i$ as
\[
\mathbf{c}^i(u,v)
    = \max_{j \in \mathbb{I}_i}\mathbf{c}^{ij}(u,v),
\]
which assigns each pixel $(u,v)$ in $I_i$ the highest correspondence confidence across all its neighboring views $\mathbb{I}_i$. The final aggregated confidence map $\mathbf{c}^{i}$ (top-right) is used to sample points. The visualization uses the \textit{treehill} scene from Mip-NeRF360~\cite{barron_mip-nerf_2022}.

Increasing the number of neighbors yields more correspondences, but this approach quickly leads to diminishing returns because different neighbors often match largely overlapping regions. In contrast, initialization time grows roughly linearly with the number of neighbors. We therefore obtain a more efficient trade-off by sampling more reference views while restricting each reference image to its 2 nearest neighbors. This provides broad scene coverage without unnecessary computational overhead.

\section{Gaussians motion through optimization}
\label{sec:sup:Gaus_motion}
We provide videos in the supplementary material (folder \texttt{sec\_D\_gaussians\_motion/}) that visualize how Gaussian 
parameters evolve during optimization. 
This experiment highlights that, thanks to our more accurate initialization, EDGS begins much closer to the final solution, 
leading to substantially smaller parameter updates and shorter optimization trajectories. 
To visualize splat motion, we record the positions and colors of all Gaussians at every iteration. 
After training, we identify the final set of Gaussians and reconstruct their trajectories by tracing their position and color 
histories backward through the optimization. 
For 3DGS, which performs densification, we additionally follow each split or cloned Gaussian back to its corresponding 
``parent'' in order to maintain consistent trajectories from the first iteration; EDGS does not require this step. 
We illustrate this analysis on the \textit{flowers} and \textit{stump} scenes from the Mip-NeRF360 dataset, 
where visualizations confirm that our splats undergo far fewer adjustments and reach high-quality reconstructions much earlier in training.

\section{Initialization with denser COLMAP}
\label{sec:supp_init_baselines}

To further test whether EDGS' improvements could be attributed to simply starting from a \emph{stronger} or \emph{denser} prior, we include additional initialization experiments for completeness. The goal is to evaluate whether ``more Gaussians'' or alternative priors can match EDGS when densification is removed. In~\cref{tab:rebuttal}, we evaluate a naïve ``denser COLMAP'' baseline by duplicating the COLMAP points $10\times$ and $50\times$ and adding noise $\epsilon\sim\mathcal{N}(0,\sigma)$ with $\sigma\in\{10^{-3},10^{-1}\}$ to the 3D coordinates to avoid exact overlap. This substantially increases training time but does not improve reconstruction quality, showing that ``more Gaussians'' alone is insufficient; EDGS' gains come from a geometrically accurate initialization rather than raw point count.

\vspace{-1pt}
\begin{table}[ht]
    \centering  
    \adjustbox{max width=0.47\textwidth}{\begin{tabular}{l  @{\hskip 4mm} c c c c c l l }
                     \textbf{Mip-NeRF 360} & \multicolumn{1}{c}{\begin{tabular}[c]{@{}c@{}}\textbf{Init}\\ \textbf{Duplication}\end{tabular}} & $\bm{\sigma}$ &\textbf{SSIM} $\uparrow$ & \textbf{PSNR} $\uparrow$ & \textbf{LPIPS} $\downarrow$ & \multicolumn{1}{c}{\begin{tabular}[c]{@{}c@{}}\textbf{Train}\\ \textbf{time}\end{tabular}} &  \multicolumn{1}{c}{\begin{tabular}[c]{@{}c@{}}\textbf{\#G} \\ ($10^6$)\end{tabular}} \\
        \toprule

              & $\times1$ &  - & 0.816  & 27.49  & 0.215 & 26 m    & 2.8 \\

          & $\times10$ &  $10^{-1}$ & 0.815 & 27.25 & 0.209 & 31 m & 3.0   \\
          & $\times10$ &  $10^{-3}$ & 0.815  & 27.22  & 0.211 & 32 m    & 3.1 \\

         & $\times50$ &  $10^{-1}$ & 0.812 & 27.10 & 0.208 & 44 m & 3.4   \\
          \multirow{-5}{*}{ 3DGS [31]} & $\times50$ &  $10^{-3}$ & 0.814 & 27.07 & 0.209 & 47 m & 3.9   \\
     
        \bottomrule
    \end{tabular}}
    \vspace{-4pt}
    \caption{ Adding more Gaussians to the COLMAP initialization is not enough.
    }
    
    \label{tab:rebuttal}
\end{table}
\vspace{-2pt}

\section{Additional visual results}
\label{sec:sup:visual}

\begin{figure*}[th]
   \centering
   \includegraphics[width=1.0\textwidth]{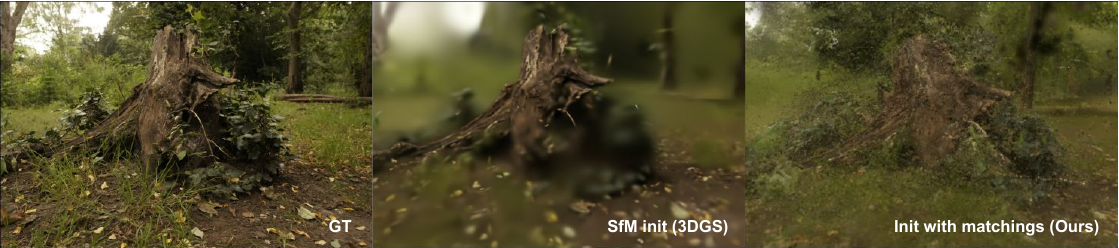}
   \caption{Visual comparison of initialization methods on the \textit{stump} scene from the Mip-NeRF360 dataset~\cite{barron_mip-nerf_2022}. The left image represents the ground truth. The middle image shows the traditional 3DGS approach initialization with Structure-from-Motion (SfM)~\cite{schonberger2016structure}. The right image illustrates initialization with our method using matchings. Despite a noisy appearance at initialization, our model can jointly optimize all the Gaussians and achieve better reconstruction quality. }
  \label{fig:inits}
  \centering
\end{figure*}

\paragraph{Initializations}
In~\cref{fig:inits}, we compare our initialization against the standard SfM-based initialization used in 3DGS. The latter produces sparse and uneven point clouds, often leaving large background regions underrepresented. In contrast, our approach initializes splats densely across the entire scene. Although the resulting initialization may appear noisy, the optimization quickly suppresses erroneous splats and retains only those consistent with target views. This dense starting point ensures that all regions receive early supervision, avoiding the multiple densification rounds required by the standard 3DGS pipeline to reach a comparable level of coverage and reconstruction quality.

\begin{figure}[!ht]
  \centering
   \includegraphics[width=1\columnwidth]{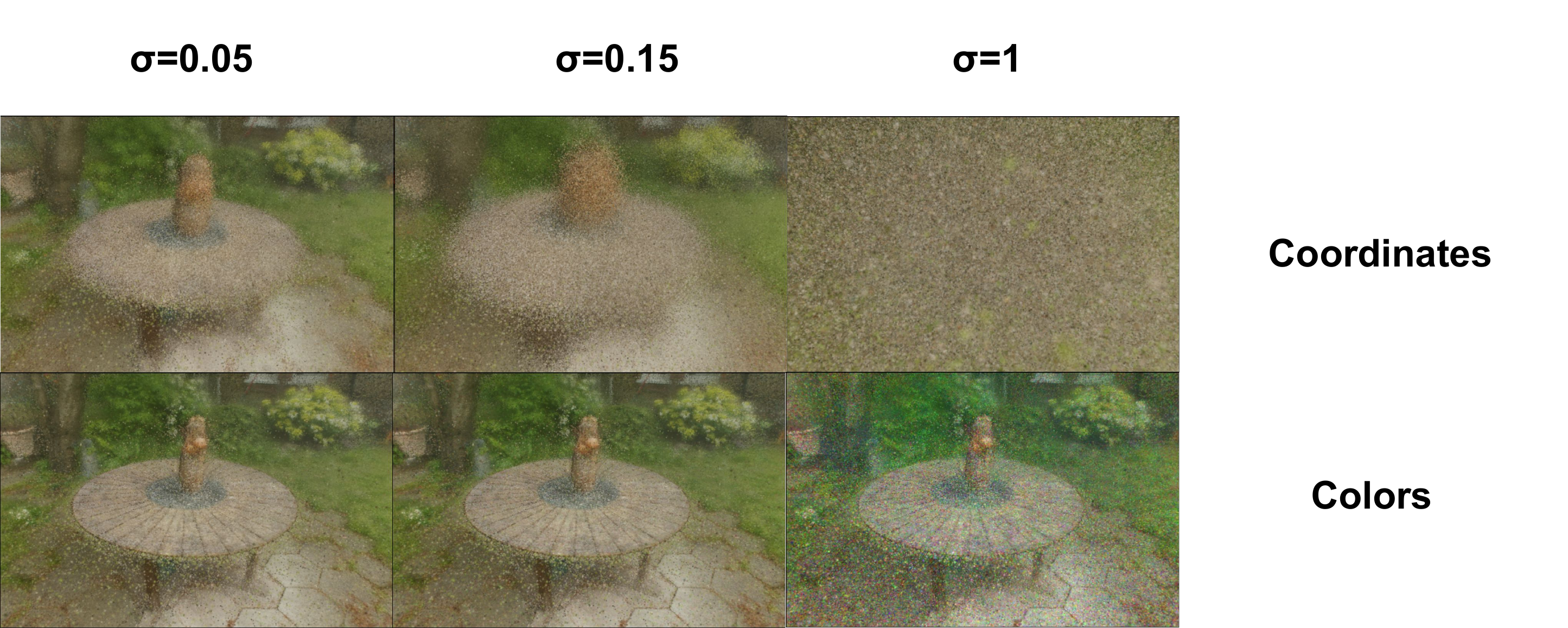}
   \caption{
The impact of noise on initialization quality. The first row shows the effect of adding noise to coordinates, while the bottom row demonstrates the effect of adding noise to color values.
   }
  \label{fig:ablation_noise_sup}
\end{figure}

\paragraph{Robustness to noise in initialization}
In~\cref{fig:ablation_noise_sup}, we visualize the effect of adding synthetic noise to our initialization. This complements the quantitative robustness analysis in the main manuscript and provides an intuitive sense of how strongly the initialization must be perturbed before reconstruction quality degrades. As the figure shows, only substantial corruption produces visibly degraded initial splat configurations. Even for noticeable noise levels such as $\sigma = 0.05$ (see the visualizations in~\cref{fig:ablation_noise_sup} and the corresponding scores in~\cref{fig:ablation_noise}), EDGS reliably recovers accurate reconstructions. Only very large perturbations, typically $\sigma > 0.15$, lead to significant degradation in performance.

\begin{figure*}[ht]
  \centering
   \includegraphics[width=1\textwidth]{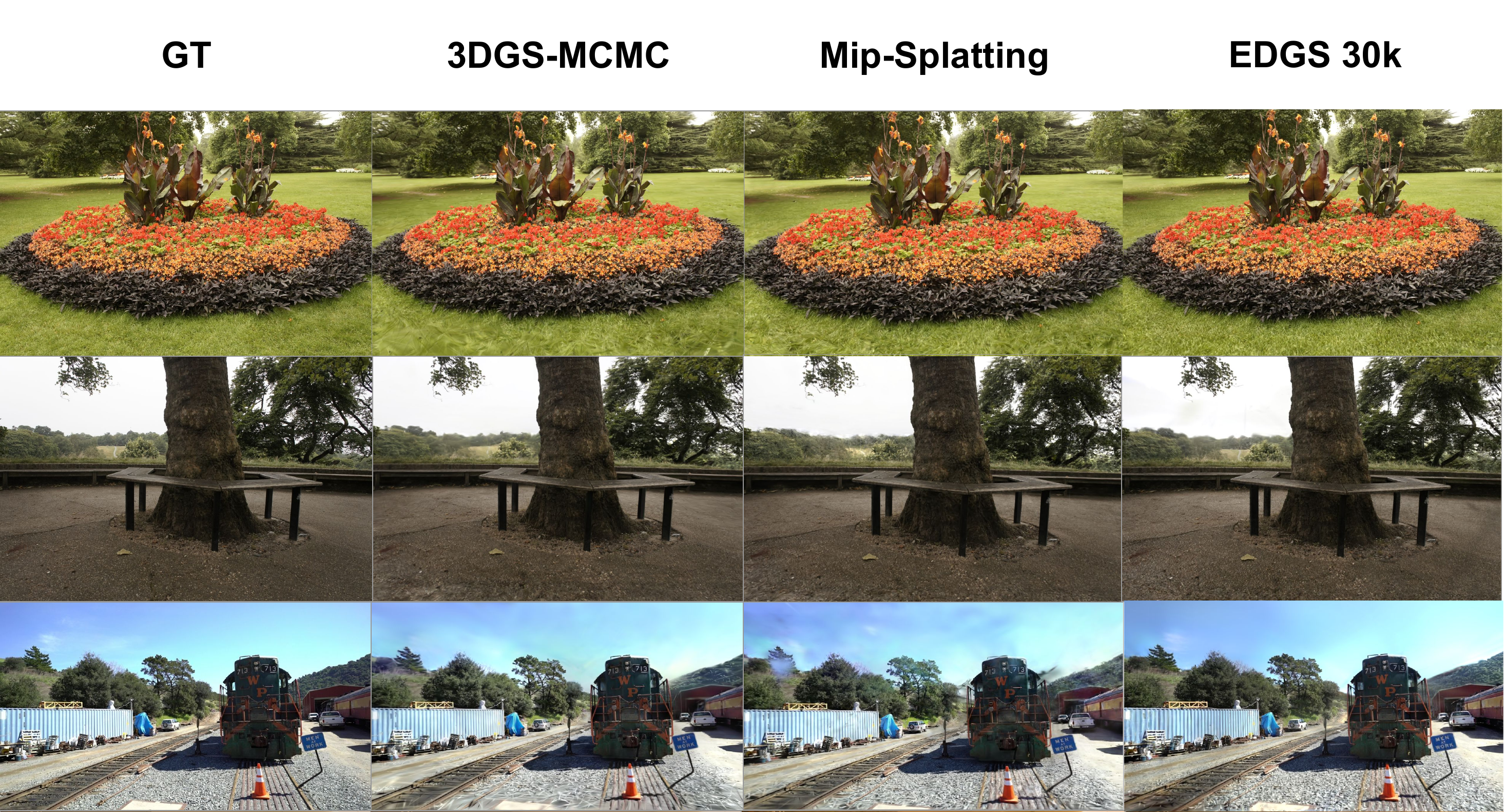}
   \caption{Additional qualitative results are presented for the scenes \textit{treehill}, \textit{flowers} and \textit{train}. For clarity, areas of interest have been zoomed in~\cref{fig:qualitative_comparison}. These results are best viewed digitally for optimal detail. }
  \label{fig:sup_visuals1}
\end{figure*}

\begin{figure*}[ht]
  \centering
   \includegraphics[width=1\textwidth]{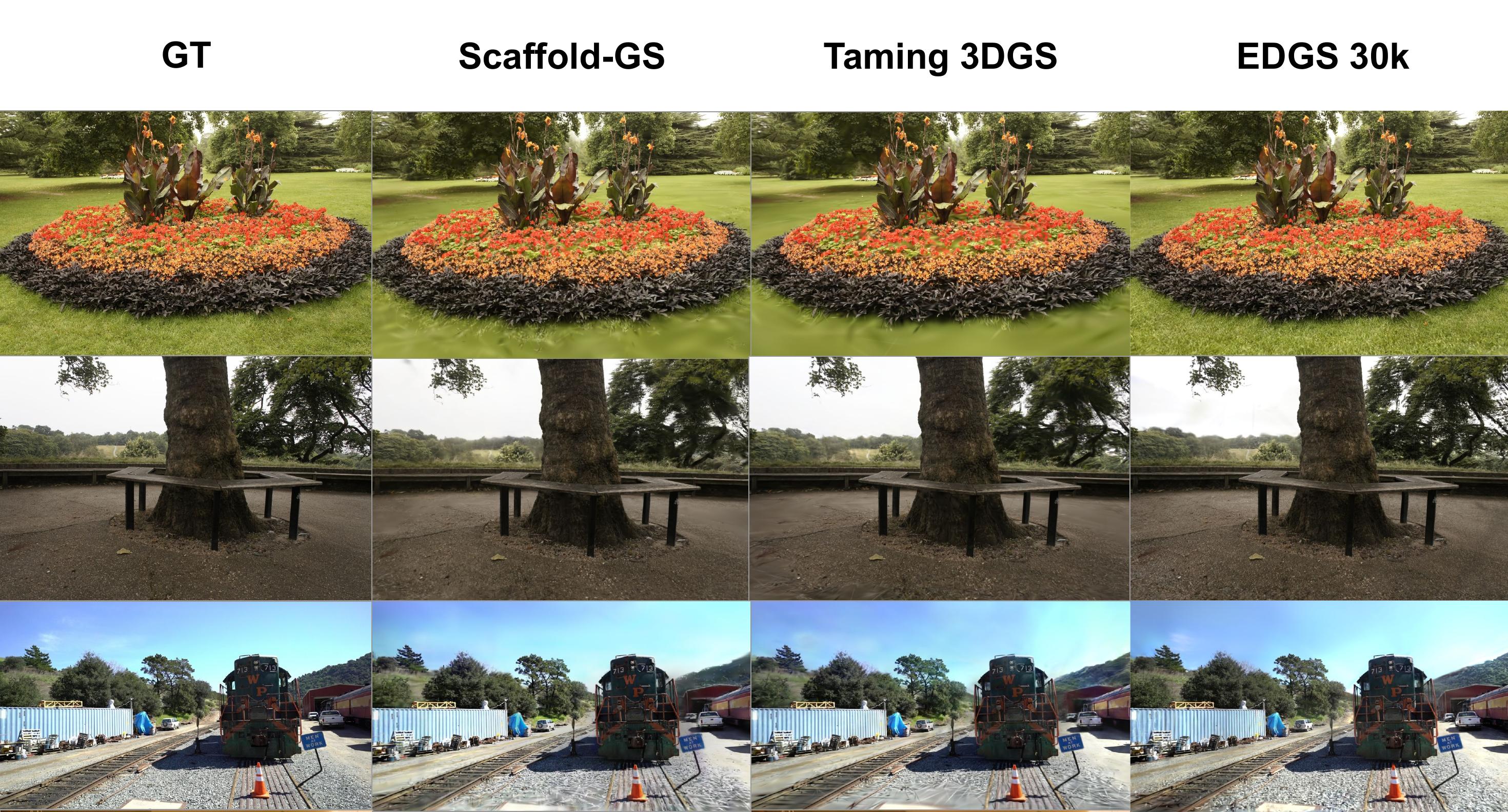}
   \caption{Additional qualitative results are presented for the scenes \textit{treehill}, \textit{flowers} and \textit{train}. For clarity, areas of interest have been zoomed in~\cref{fig:qualitative_comparison}. These results are best viewed digitally for optimal detail. }
   \label{fig:sup_visuals2}
\end{figure*}

\begin{figure*}[ht]
  \centering
   \includegraphics[width=1\textwidth]{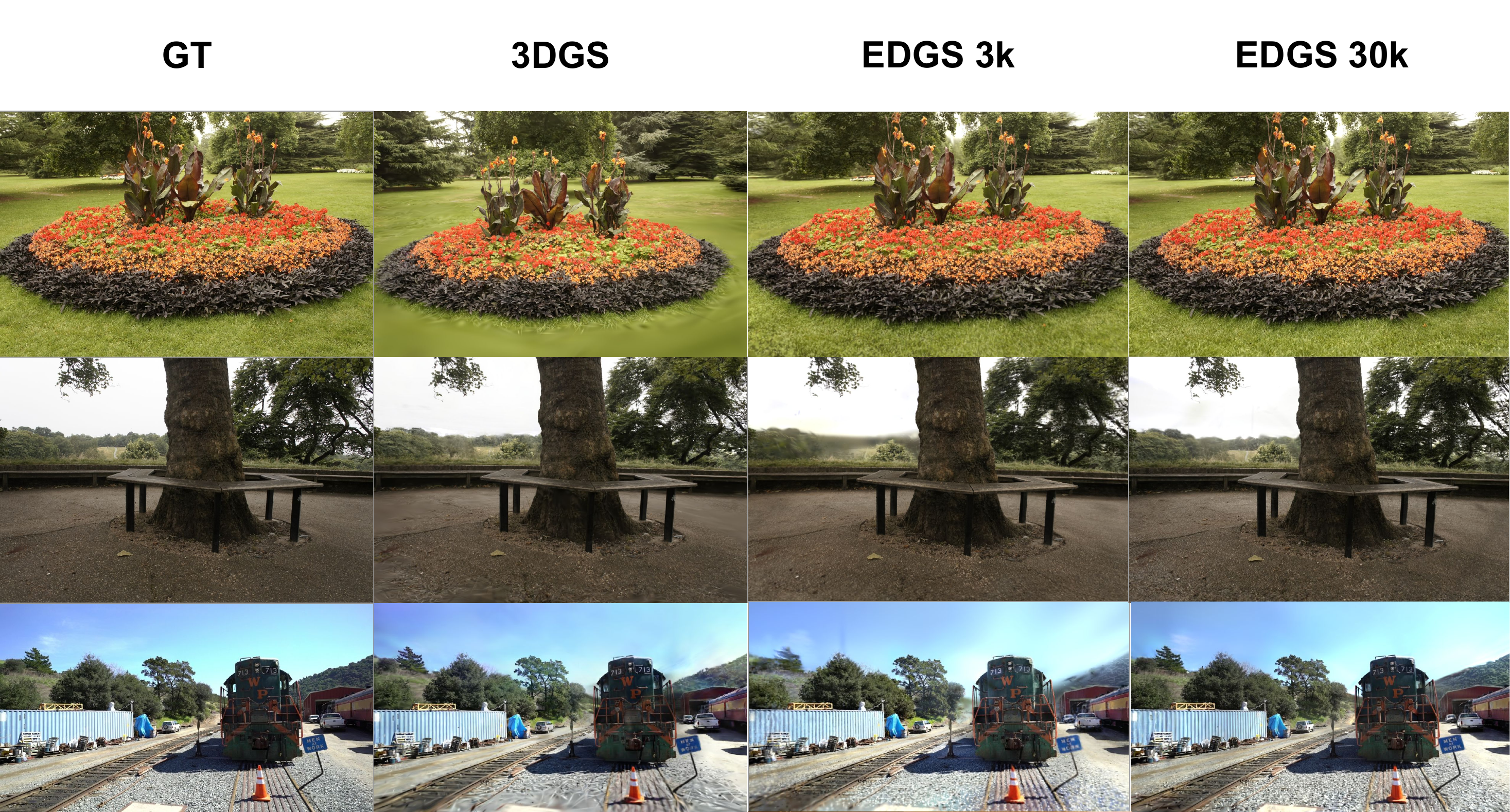}
   \caption{Additional qualitative results are presented for the scenes \textit{treehill}, \textit{flowers} and \textit{train}. For clarity, areas of interest have been zoomed in~\cref{fig:qualitative_comparison}. These results are best viewed digitally for optimal detail. }
  \label{fig:sup_visuals3}
\end{figure*}

\paragraph{Qualitative comparison}
Full-resolution versions of the renders shown in the main paper are provided in~\cref{fig:sup_visuals1,fig:sup_visuals2,fig:sup_visuals3}. For clearer comparison in~\cref{fig:sup_visuals3}, we also include renderings for the original 3DGS method.

\paragraph{Video results for front-facing scenes} Our method is also compatible with front-facing scenes. We use sequences with primarily forward-facing camera orientations, with little variation of viewpoint. From each scene, we extract 24 frames and first process them with COLMAP to recover camera intrinsics and extrinsics. We compare the original 3DGS pipeline with EDGS on the same data. For each scene, we visualize three stages of our approach: the initial splat placement, the intermediate fitting stage, and the final rendering from multiple viewpoints. We use 24 reference frames and 10,000 correspondences per reference. The results can be found in the folder \texttt{sec\_E\_front\_facing\_scenes/} of the supplementary material.

\paragraph{Video results for synthetic data}
We also visualize performance on the Synthetic dataset~\cite{mildenhall2021nerf}. We compare the original 3DGS pipeline with EDGS on the same data. Both methods are allocated equal runtime budgets, allowing for a direct comparison at matching optimization time. In both settings, we align the optimization timeline to ensure comparability. EDGS consistently achieves high-quality reconstructions faster than 3DGS, thanks to its rich and dense initialization, which provides the necessary detail from the very beginning. See folder \texttt{sec\_E\_synthetic\_scenes/} in the supplementary material.

\section{Per-scene results}
\label{sec:sup:per_scene_results}

We provide a more detailed evaluation of \textit{EDGS + 3DGS} from~\cref{tab:merged_comparison}, \textit{EDGS +3DGS 5K} from~\cref{tab:efficciency_comparison} and \textit{3DGS MCMC with EDGS init} from~\cref{tab:ours_plus_adc}. We include per-scene scores for these models in~\cref{tab:sup_SSIM360,tab:sup_PSNR360,tab:sup_LPIPS360,tab:sup_GS360,tab:sup_time360,tab:sup_TANK}.  Note that densification is disabled for the first three models; for the final model, we begin with fewer points and apply the adaptive densification strategy from 3DGS‑MCMC.

\section{Notation}
\label{sec:sup:notation}
To simplify the understanding of the paper, we include a table of notation~\cref{tab:notation}. This table provides a concise summary of the key symbols and terms used throughout the paper, along with their definitions. 


\begin{table}[ht!]
\centering
\adjustbox{max width=\columnwidth}{
\begin{tabular}{l|cc|cc}
\toprule
EDGS + 3DGS                  & Truck          & Train          & Dr Johnson     & Playroom        \\ \midrule

SSIM &  0.898 & 0.837          & 0.900          & 0.907        \\
PSNR &  26.16 & 22.39          & 29.50          & 30.12       \\
LPIPS  &  0.091 & 0.172          & 0.233          & 0.213        \\
\# Gaussians &  1.6 & 1.1          & 1.5         & 1.7        \\
Time in minutes &  25 & 21          & 30          & 30        \\

\bottomrule
\end{tabular}}
\caption{Per-scene quantitative results on the Tanks \& Temples and Deep Blending subsets.}
\label{tab:sup_TANK}
\end{table}
\begin{table*}[th!]
\centering
\adjustbox{max width=\textwidth}{
\begin{tabular}{ll ccccc|cccc}
\toprule
                  & bicycle        & flowers        & garden         & stump          & treehill       & room           & counter        & kitchen         & bonsai         \\
\midrule
EDGS + 3DGS 5K             & 0.760 &	0.619 &	0.853 &	0.789 &	0.654 &	0.941 &	0.912 &	0.943	& 0.950\\
EDGS + 3DGS            & 0.792 &	0.641 &	0.876 &	0.783 &	0.655 &	0.954 &	0.931 &	0.955	& 0.962\\
3DGS-MCMC + EDGS Init  & 0.815 &	0.664 &	0.891 &	0.815 &	0.666 &	0.941 &	0.928 &	0.943	& 0.959\\
\bottomrule
\end{tabular}}
\caption{Per-scene quantitative results (SSIM) on the Mip-NeRF360.}
\label{tab:sup_SSIM360}
\end{table*}

\begin{table*}[th!]
\centering
\adjustbox{max width=\textwidth}{
\begin{tabular}{l ccccc|cccc}
\toprule
                 & bicycle         & flowers        & garden          & stump           & treehill       & room           & counter        & kitchen        & bonsai         \\ 
\midrule

EDGS + 3DGS 5K              & 24.58	& 21.37	& 26.75	& 26.78	& 22.32 &	30.63	& 27.91 &	30.06 &	30.96 \\
EDGS + 3DGS               & 25.39	& 21.57	& 27.67	& 26.67	& 22.47 &	32.87	& 29.62 &	32.99 &	32.96 \\
3DGS-MCMC + EDGS Init  & 26.14   & 21.85	& 28.39	& 27.42	& 22.8 &	32.47	& 29.62 &	32.47 &	33.41 \\
\bottomrule
\end{tabular}}
\caption{Per-scene quantitative results (PSNR) on the Mip-NeRF360.}
\label{tab:sup_PSNR360}
\end{table*}

\begin{table*}[th!]

\centering
\adjustbox{max width=\textwidth}{
\begin{tabular}{l ccccc|cccc}
\toprule
                    & bicycle        & flowers        & garden         & stump          & treehill       & room           & counter        & kitchen        & bonsai         \\ \midrule

EDGS + 3DGS 5K             & 0.203	& 0.310	& 0.120 &	0.203	& 0.278	& 0.110	& 0.113	& 0.073  &	0.085\\
EDGS + 3DGS             & 0.161	& 0.267	& 0.095 &	0.192	& 0.252	& 0.089	& 0.088	& 0.059  &	0.070\\
3DGS-MCMC + EDGS Init  & 0.145	& 0.242	& 0.084 &	0.165	& 0.239	& 0.165	& 0.149	& 0.1  &	0.144\\

\bottomrule
\end{tabular}}
\caption{Per-scene quantitative results (LPIPS) on the Mip-NeRF360.}
\label{tab:sup_LPIPS360}
\end{table*}

\begin{table*}[th!]

\centering
\adjustbox{max width=\textwidth}{
\begin{tabular}{l ccccc|cccc}
\toprule
                   & bicycle        & flowers        & garden         & stump          & treehill       & room           & counter        & kitchen        & bonsai         \\ \midrule

EDGS + 3DGS 5K         & 2.8 &	2.5	& 2.8	& 1.9	& 2.2 &	2.9 &	2.8 &	3.0	& 2.6 \\
EDGS + 3DGS       & 2.3 &	2.2	& 2.5	& 1.8	& 2.0 &	1.3 &	1.7 &	1.7	& 1.3 \\
3DGS-MCMC + EDGS Init  & 5.9	&3.7	&5.2	&4.8	&3.6	&1.5	&1.3	&1.8	&1.4\\

\bottomrule
\end{tabular}}
\caption{Per-scene quantitative results (millions of Gaussians $\#G$ ) on the Mip-NeRF360.}
\label{tab:sup_GS360}
\end{table*}

\begin{table*}[th!]

\centering
\adjustbox{max width=\textwidth}{
\begin{tabular}{l ccccc|cccc}
\toprule
                   & bicycle        & flowers        & garden         & stump          & treehill       & room           & counter        & kitchen        & bonsai         \\ \midrule

EDGS + 3DGS 5K           & 9	& 8	& 9 &	7 &	8 &	7 &	7 &	8 & 7 \\
EDGS + 3DGS           & 31	& 30	& 34 &	27 &	31 &	20 &	23 &	25 & 22 \\
3DGS-MCMC + EDGS Init  & 32	& 22	& 29 &	26 &	21 &	12 &	12 &	14 & 12 \\
\bottomrule
\end{tabular}}
\caption{Per-scene quantitative results (time in minutes) on the Mip-NeRF360.}
\label{tab:sup_time360}
\end{table*}

\begin{table*}[ht]
\centering
\caption{Notation}
\label{tab:notation}
\begin{tabular}{p{3cm}p{10cm}}
\toprule
\textbf{Notation} & \textbf{Description} \\
\midrule
$\mathbb{G}$ & Set of 3D Gaussians representing the scene \\
$\bm{g}_i$ & $i$‑th Gaussian in $\mathbb{G}$, with parameters $\{\bm{g}^x_i,\bm{\Sigma}_i,\bm{g}^c_i,\bm{g}^{\alpha}_i\}$ \\
$\bm{g}^x_i\in\mathbb{R}^3$ & 3D center of Gaussian $i$ \\
$\bm{\Sigma}_i\in\mathbb{R}^7$ & Encoded covariance (shape) of Gaussian $i$ \\
$\bm{g}^c_i\in\mathbb{R}^3$ & RGB color of Gaussian $i$ \\
$\bm{g}^{\alpha}_i\in\mathbb{R}$ & Opacity of Gaussian $i$ \\
$p$ & Pixel location in the rendered image \\
$C(p)$ & Rendered color at pixel $p$ \\
$(\bm{p}' - \bm{g}^x_i)$ & Shortest distance between the pixel projection line and $\bm{g}^x_i$ \\
$\bm{\sigma}_i(p)$ & Contribution of Gaussian $i$ to pixel $p$ \\
$\bm{R}_i,\;\bm{S}_i$ & Rotation and scaling for $\bm{\Sigma}_i$  \\
$I^i$ & Reference image \\
$\mathbb{I} = \{I^j\}$ & Set of neighboring images for $I^i$ \\
$\bm{P}^i\in\mathbb{R}^{3\times4}$ & Projection matrix of camera $i$ \\
$\mathcal{M}$ & Pretrained dense matching network \\
$\mathcal{W}^{i\rightarrow j}\in\mathbb{R}^{2\times H\times W}$ & Warp field from $I^i$ to $I^j$ \\
$\mathbf{c}^{ij}\in\mathbb{R}^{H\times W}$ & Confidence of correspondences between $I^i$ and $I^j$ \\
$(u_k^i,v_k^i),(u_k^j,v_k^j)$ & Matched pixel coordinates in $I^i$ and $I^j$ \\
$\bm{g}_k^x\in\mathbb{R}^3$ & 3D position of the $k$‑th new Gaussian (via triangulation) \\
$w_k^i,w_k^j$ & Homogeneous-scale factors in projection equations \\
$\pi(\bm{P},\cdot)$ & Projection with camera matrix $\bm{P}$ \\
$\varepsilon_k^i$ & reprojection error in the reference image $I^i$ for $\bm{g}_k$ \\
$\tau_{corr}$ & Confidence threshold for sampling 2D correspondences \\
$\tau_{proj}$ & Threshold for reprojection error \\
$\mathbf{p}^{ij}_{\text{corr}}(u,v)$ & Uniform sampling distribution over $\{(u_i^k,v_i^k)\mid \mathbf{c}^{ij}(u,v)>\tau_{corr} \}$ \\
$\mathbf{p}^{ij}_{\text{proj}}(u, v)$ & Uniform sampling distribution over $\{(u_i^k,v_i^k) \mid \varepsilon^{ij}_k < \tau_{\text{proj}}\}$\\
$\mathbf{p}^i(k)$ & Combined sampling distribution for image $I^i$  \\
$\mathbf{v}_1, \dots, \mathbf{v}_n \in \mathbb{R}^3$ & View directions \\
$\mathbf{Y}_k\in\mathbb{R}^{n\times16}$ & Spherical‐harmonic basis evaluated \ for $n$ views \\
$\mathbf{O}_k\in\mathbb{R}^{n\times3}$ & Observed RGB colors of splat $k$ in $n$ views\\
$\hat{\mathbf{H}}_k\in\mathbb{R}^{16\times3}$ & Fitted spherical‐harmonic coefficients \\
\bottomrule
\end{tabular}
\end{table*}

\end{document}